\documentclass[aps,prb,twocolumn,amsmath,amssymb,superscriptaddress,floatfix,longbibliography]{revtex4-1}
\usepackage{amsmath}
\usepackage{amssymb}
\usepackage{amsfonts}
\usepackage{listings}
\usepackage{enumerate}
\usepackage{latexsym}
\usepackage{bm}
\usepackage{graphicx}
\usepackage{siunitx}
\usepackage{subfigure}
\usepackage{braket}
\usepackage[pdftex,colorlinks=false]{hyperref}
\usepackage{xcolor}
\usepackage{verbatim}

\usepackage{times}

\newcommand{\beq}{\begin{equation}}
\newcommand{\eneq}{\end{equation}}
\newcommand{\bs}[1]{\boldsymbol{#1}}

\def\be{\begin{equation}}
\def\ee{\end{equation}}
\def\ba{\begin{eqnarray}}
\def\ea{\end{eqnarray}}

\def\R{{\rm Re}}
\def\Z{\mathbb{Z}}
\def\C{\mathbb{C}}

\def\beq{\begin{equation}}
\def\eeq{\end{equation}}
\def\barray{\begin{eqnarray}}
\def\earray{\end{eqnarray}}

\font\upright=cmu10 scaled\magstep1
\def\stroke{\vrule height8pt width0.4pt depth-0.1pt}

\def\Zmath{\mathbb{Z}}
\def\Qmath{\vcenter{\hbox{\upright\rlap{\rlap{Q}\kern
                   3.8pt\stroke}\phantom{Q}}}}
\def\Nmath{\vcenter{\hbox{\upright\rlap{I}\kern 1.7pt N}}}
\def\Cmath{\vcenter{\hbox{\upright\rlap{\rlap{C}\kern
                   3.8pt\stroke}\phantom{C}}}}
\def\Rmath{\vcenter{\hbox{\upright\rlap{I}\kern 1.7pt R}}}
\def\Z{\ifmmode\Zmath\else$\Zmath$\fi}
\def\Q{\ifmmode\Qmath\else$\Qmath$\fi}
\def\N{\ifmmode\Nmath\else$\Nmath$\fi}
\def\C{\ifmmode\Cmath\else$\Cmath$\fi}
\def\R{\ifmmode\Rmath\else$\Rmath$\fi}

\input{epsf}

\usepackage{bbold}

\newcounter{defcounter}
\begin{document}

\newcommand{\beginsupplement}{%
        \setcounter{table}{0}
        \renewcommand{\thetable}{S\arabic{table}}%
        \setcounter{figure}{0}
        \renewcommand{\thefigure}{S\arabic{figure}}%
        \setcounter{equation}{0}
        \renewcommand{\theequation}{S\arabic{equation}}
     }

\tolerance 10000

\newcommand{\cbl}[1]{\color{blue} #1 \color{black}}

\newcommand{\vk}{{\bf k}}

\widowpenalty10000
\clubpenalty10000

\title{Higher-Order Topology in Bismuth}

\author{
Frank~Schindler}
\address{
 Department of Physics, University of Zurich, Winterthurerstrasse 190, 8057 Zurich, Switzerland
}

\author{
Zhijun~Wang}
\address{Department of Physics, Princeton University, Princeton, New Jersey 08544, USA}

\author{
Maia~G.~Vergniory}
\address{
 Donostia International Physics Center, P. Manuel de Lardizabal 4, 20018 Donostia-San Sebastian, Spain
}
\address{
 Department of Applied Physics II, Faculty of Science and Technology,
University of the Basque Country UPV/EHU, Apartado 644, 48080 Bilbao, Spain
}
\address{
IKERBASQUE, Basque Foundation for Science, Maria Diaz de
Haro 3, 48013 Bilbao, Spain
}

\author{
Ashley~M.~Cook}
\address{
 Department of Physics, University of Zurich, Winterthurerstrasse 190, 8057 Zurich, Switzerland
}

\author{Anil~Murani}
\address{LPS, Univ. Paris-Sud, CNRS, UMR 8502, F-91405 Orsay Cedex, France}
 
\author{Shamashis~Sengupta}
\address{CSNSM, Univ. Paris-Sud, IN2P3, UMR 8609, F-91405 Orsay Cedex, France}
  
\author{Alik~Yu.~Kasumov}
\address{LPS, Univ. Paris-Sud, CNRS, UMR 8502, F-91405 Orsay Cedex, France}
\address{Institute of Microelectronics Technology and High Purity Materials, RAS, ac. Ossipyan, 6,  Chernogolovka, Moscow Region, 142432, Russia}
 
\author{Richard~Deblock}
\address{LPS, Univ. Paris-Sud, CNRS, UMR 8502, F-91405 Orsay Cedex, France}

\author{Sangjun~Jeon}
\address{Joseph Henry Laboratories and Department of Physics, Princeton University, Princeton, New Jersey 08544, USA}

\author{Ilya~Drozdov}
\address{
Condensed Matter Physics and Materials Science Department, Brookhaven National Laboratory, Upton, New York 11973, USA
}

\author{H\'el\`ene~Bouchiat}
\address{LPS, Univ. Paris-Sud, CNRS, UMR 8502, F-91405 Orsay Cedex, France}
 
\author{Sophie~Gu\'eron}
\address{LPS, Univ. Paris-Sud, CNRS, UMR 8502, F-91405 Orsay Cedex, France}

\author{Ali~Yazdani}
\address{Joseph Henry Laboratories and Department of Physics, Princeton University, Princeton, New Jersey 08544, USA}

\author{
B.~Andrei~Bernevig}
\address{Joseph Henry Laboratories and Department of Physics, Princeton University, Princeton, New Jersey 08544, USA}

\author{
Titus~Neupert}
\address{
 Department of Physics, University of Zurich, Winterthurerstrasse 190, 8057 Zurich, Switzerland
}

\begin{abstract}
The mathematical field of topology has become a framework to describe the low-energy electronic structure of crystalline solids.
A typical feature of a bulk insulating three-dimensional topological crystal are conducting two-dimensional surface states. This constitutes the topological bulk-boundary correspondence. 
 Here, we establish that the electronic structure of bismuth, an element consistently described as bulk topologically trivial, is in fact
 topological and follows a generalized bulk-boundary correspondence of higher-order:
 not the surfaces of the crystal, but its hinges host topologically protected conducting modes.  
These hinge modes are protected against localization by time-reversal symmetry locally, and  globally by the three-fold rotational symmetry and inversion symmetry of the bismuth crystal. 
We support our claim theoretically and experimentally. Our theoretical analysis is based on symmetry arguments, topological indices, first-principle calculations, and the recently introduced framework of topological quantum chemistry. 
We provide supporting evidence from two complementary experimental techniques. With scanning-tunneling spectroscopy, we probe the unique signatures of the rotational symmetry of the one-dimensional states located at step edges of the crystal surface. With Josephson interferometry, we demonstrate their universal topological contribution to the electronic transport.
Our work establishes bismuth as a higher-order topological insulator. 
\end{abstract}

\date{\today}

\maketitle

\begin{figure*}[t]
\begin{center}
\includegraphics[width=0.9\textwidth]{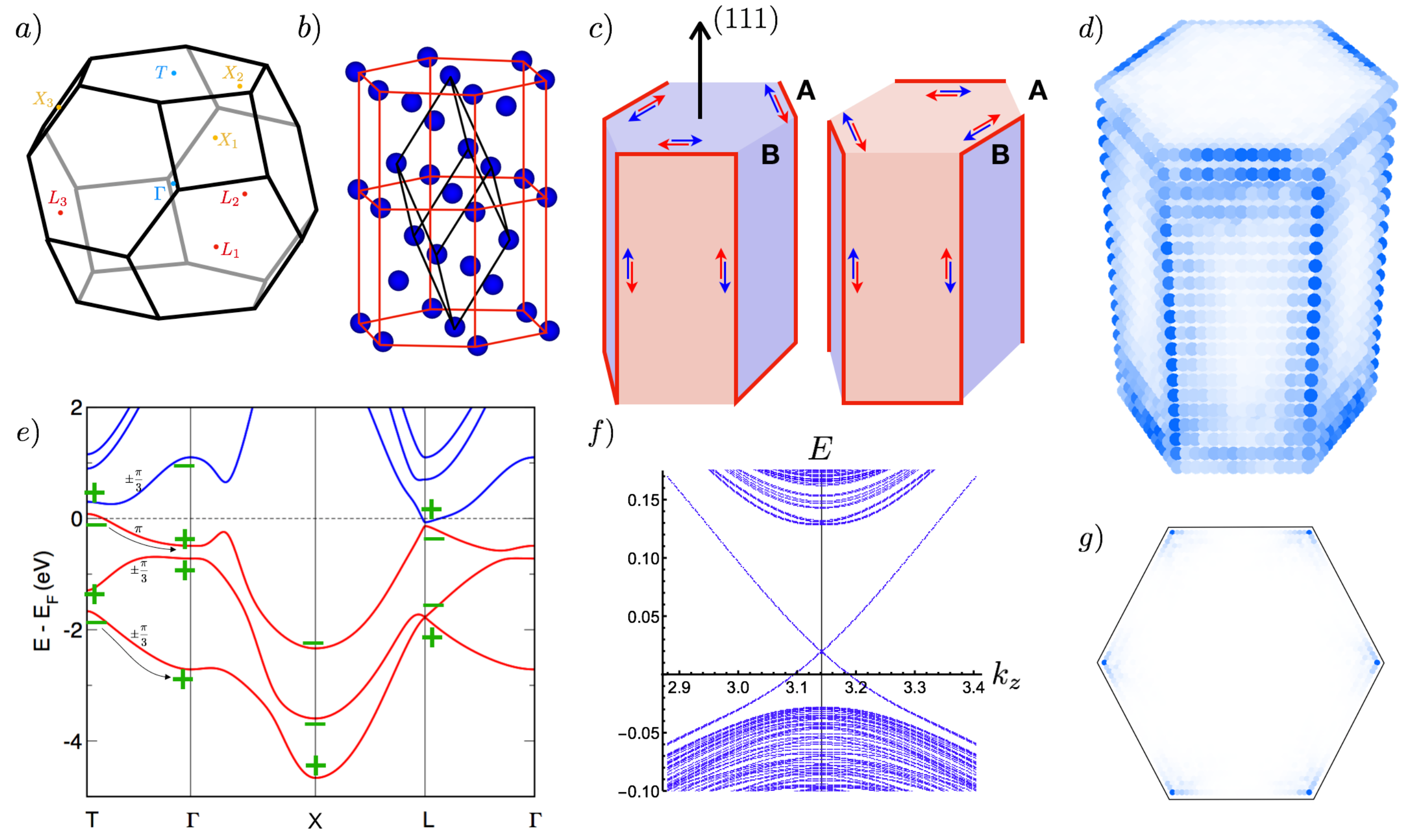}
\caption{
Electronic structure of a HOTI with $\hat{C}_3$ and $\hat{I}$. 
a) Brillouin zone with TRS points that are used to evaluate the topological indices in Eq.~\eqref{eq: topo index}.
b) Unit cell of the crystal structure of bismuth, which has $\hat{C}_3$ and inversion-symmetry. There are six inequivalent sites in the conventional (hexagonal) unit cell, which is shown in red. Black lines delineate the primitive unit cell (rhombohedral), which has only 2 inequivalent atoms.
c) Schematic of the hinge states of a hexagonally-shaped HOTI oriented along the trigonal [111] axis, with $\hat{C}_3$ and inversion-symmetry (e.g., bismuth). Note that a prism with triangular rather than hexagonal cross-section would would not respect inversion symmetry. All edges of the hexagonal cross-section are along bisectrix axes.
Red lines represent a single one-dimensional Kramers pair of gapless protected modes. In the Dirac picture of a HOTI surface, red and blue surfaces correspond to opposite signs of the unique TRS surface mass terms.
d) Localized hinge modes of the minimal tight-binding model of a HOTI with the same topology and symmetries as bismuth, as defined in the Supplementary Information. The model is solved on the hexagon geometry described in (c) with open boundary conditions in all directions. Plotted is the sum of the absolute squares of the eigenstates that lie in the bulk and surface gap. Note that while the tight-binding model considered has the same topology as bismuth, it lacks its metallic surface states which are not protected by $\hat{C}_3$ and inversion symmetry.
e) Band structure of bismuth with inversion eigenvalues (green) and $\hat{C}_3$ eigenvalues on the $\Gamma$--$T$ line (black). Since valence bands (red) and conduction bands (blue) are not degenerate anywhere in momentum space, their topological indices, Eq.~\eqref{eq: topo index}, are well defined despite the appearance of a small electron and hole pocket. Black arrows indicate the two valence bands contributing to the $\hat{C}_3$-eigenvalue-graded band inversion.
f) Spectrum of the same model solved on a nanowire with hexagonal cross-section and periodic boundary conditions in the trigonal $z$ direction ([111] direction).
Only a portion of the spectrum at small momentum deviations from the $T$ point $k_z = \pi$ is shown. Six Kramers pairs of hinge modes traverse the surface and bulk gap. Consult Fig.~\ref{fig: tightbinding}~c) in the Supplementary Information for a zoomed-out version showing the spectrum for all momenta.
g) Localization of these topologically protected hinge modes in the $x$-$y$-plane. 
}
\label{fig: rod}
\end{center}
\end{figure*}

%%%%%%%%%%%%%%%%%%%%%%%%%%%%%%%%%%%%%%%%%%%%%%%%%%%%

Electronic insulators are characterized by an energy gap between valence and conduction bands.
Two insulators are classified as topologically equivalent, if they can be deformed into one another without closing this gap, while certain symmetries are respected.
If time-reversal symmetry (TRS) is respected in this deformation process, three-dimensional (3D) insulators have been shown to fall into two disconnected topological classes: trivial and nontrivial~\cite{Kane07,Moore07,Hsieh09, Roy2009,FuKane2007,Xia2009,Chen2009}. The latter are called topological insulators (TIs).
What makes this mathematical classification highly relevant experimentally is the so-called bulk-boundary correspondence of TIs: the two-dimensional (2D) surface of a 3D TI hosts conducting states, with the dispersion of a non-degenerate Dirac cone, which cannot be gapped or localized without breaking TRS (or inducing interacting instabilities such as superconductivity or topological order).
When, in addition, the spatial symmetries of the crystal are preserved during this deformation process, such as mirrors or rotations, one speaks of topological crystalline insulators~\cite{Fu11, Hsieh12, Dziawa12, Mong2010, Tanaka2012, Xu2012a}. The additional symmetries were argued to stabilize multiple Dirac cones on surfaces that are invariant under both TRS and the protecting spatial symmetry. 

More recently, topological crystalline insulators were generalized to also include \textit{higher-order topological insulators} (HOTIs)~\cite{Benalcazar16,SchindlerHOTI}, in which gapless topological states protected by spatial symmetries appear at corners or hinges, while the edges and surfaces are gapped in 2D and 3D systems, respectively. HOTIs thus generalize the topological bulk-boundary correspondence. While the topological protection of point-like corner modes requires some spectral symmetry, one-dimensional gapless hinge modes mediate a spectral flow~\cite{SchindlerHOTI,Song17,Benalcazar17,Langbehn17} between valence and conduction band of the bulk insulator, akin to quantum Hall~\cite{Klitzing80, Thouless82, Avron1985, Niu1985} or quantum spin Hall edge modes~\cite{Kane05a, Kane05b, Qi2006, Bernevig2006a, Bernevig2006b, Koenig07, Roth2009, Wu2006}. Therefore, they can be expected to appear more generically in actual crystalline materials. Several works studied the classification of \mbox{HOTIs}~\cite{Song17,Langbehn17,Xu17,Shapourian17,Lin17,Ezawa18,Khalaf18,Geier18}, for example in the presence of two-fold spatial symmetries~\cite{Song17} or $\hat{C}_n$ rotational symmetries~\cite{Fang17}.

Various topological aspects of the electronic structure of bismuth have been studied experimentally in the past.
This revealed intriguing features such as one-dimensional topological modes localized along step edges on the surface of bismuth~\cite{Drozdov2014}, conducting hinge channels on bismuth nanowires\cite{Li2014-2,Murani17}, 
quasi-one-dimensional metallic states on the bismuth (114) surface\cite{Wells09}, and a quantum spin Hall effect in 2D bismuth bilayers~\cite{Murakami06,Takayama15} and bismuthene on silicon carbide~\cite{Reis2017}.
Here, we show, based on band representations and the theory of topological quantum chemistry~\cite{Bradlyn17,Vergniory17,elcoro2017double,Cano17-2,Bradlyn17-2,Cano17}, that bismuth is in fact a HOTI. This finding provides a unified theoretical origin for all these previous experimental observations. The crystal symmetries that protect the topology of bismuth, $\hat{C}_3$ rotation and inversion, establish a new class of HOTIs not discussed in previous works\cite{Benalcazar16,SchindlerHOTI,Song17,Benalcazar17,Langbehn17,Xu17,Shapourian17,Lin17,Khalaf18,Ezawa18,Ezawa218,Geier18}. 
We support our theoretical analysis with experimental data using two complementary techniques: scanning tunneling microscopy/spectroscopy (STM/STS) on surface step edges and Josephson interferometry on bismuth nanowires.

Another 3D material that hosts one-dimensional modes on its surface is the topological crystalline insulator tin telluride. For one, strained tin telluride is proposed to become a HOTI~\cite{SchindlerHOTI}. In addition, tin telluride has been experimentally shown to feature one-dimensional flatband modes localized at surface step edges~\cite{Sessi2016}. The latter appear together with the Dirac cone topological surface states and are thus distinct from the hinge modes of a HOTI.

Note that while there are analog experimental realizations of 2D second-order topological insulators via electrical circuits\cite{Imhof17}, as well as phononic\cite{Garcia18} and photonic\cite{Peterson18} systems, the present work provides the first instance of a realization of the concept in the electronic structure of a crystal. At the same time, it is the first experimental confirmation of a 3D HOTI, regardless of the platform.

%We first develop the theoretical arguments and subsequently present the two types of experimental evidence for topological hinge states in bismuth.

\textit{Bulk topology} --- 
Fu and Kane~\cite{FuKane2007} gave a simple topological index for a 3D TI in the presence of inversion symmetry $\hat{I}$: One multiplies the inversion eigenvalues (which are $\pm1$) of all Kramers pairs of occupied bands at all time-reversal symmetric momenta (TRIMs) in the Brillouin zone. If this product is $-1$ ($+1$), the insulator is topological (trivial). In the topological case, one says the material has a band inversion.
Note that when we evaluate this index for bismuth, we obtain $+1$, in accordance with the well known result that the band structure of bismuth is topologically trivial from a first-order perspective\cite{Teo08}. A sample of bismuth thus does not have topologically protected gapless surface states. However, this is not due to bismuth not displaying a band inversion: in fact, we will show that there are two band inversions, the presence of which is not captured by the first-order index, which is only sensitive to the parity of band inversions. We first extend this index to HOTIs with TRS, $\hat{C}_3$ rotation, and inversion symmetry $\hat{I}$. Note that we consider a $\hat{C}_3$ rotational symmetry with axis that is given by the line connecting the TRIMs $\Gamma$ and $T$ [consult Fig.~\ref{fig: rod}~a) for a representation of the Brillouin zone]. For spin-$1/2$ particles, $\hat{C}_3$ has eigenvalues $-1$ and $\mathrm{exp}(\pm\mathrm{i}\pi/3)$, where a subspace with $-1$ eigenvalue is closed under TRS, while TRS maps the $\mathrm{exp}(+\mathrm{i}\pi/3)$ subspace to the $\mathrm{exp}(-\mathrm{i}\pi/3)$ one and vice versa. We can thus define a band inversion separately in the occupied band subspaces of an insulator with $\hat{C}_3$ eigenvalues $-1$ and $\mathrm{exp}(\pm\mathrm{i}\pi/3)$. To do so, observe that of the eight TRIMs, two are invariant under $\hat{C}_3$ ($\Gamma$ and $T$), while two groups of three TRIMS transform into each other under $\hat{C}_3$ (call them $X_i$ and $L_i$, $i=1,2,3$). Denote by $\nu_Y=\prod_{i\in\mathrm{occ}}\xi_{i,Y}$ the product over all inversion eigenvalues $\xi_{i,Y}=\pm1$ of the occupied bands Kramers pairs at the TRIM $Y\in \{\Gamma, T, X_i,L_i\}$. At $\Gamma$ and $T$ we further define $\nu_Y^{(\pi)}$ and $\nu_Y^{(\pm\pi/3)}$, where the product is restricted to the Kramers pairs with $\hat{C}_3$ eigenvalues $-1$ and $\mathrm{exp}(\pm\mathrm{i}\pi/3)$, respectively, such that $\nu_Y=\nu_Y^{(\pi)}\nu_Y^{(\pm\pi/3)}$ for $Y=\Gamma, T$. By $\hat{C}_3$ symmetry 
$\nu_{X_1}=\nu_{X_2}=\nu_{X_3}$ and 
$\nu_{L_1}=\nu_{L_2}=\nu_{L_3}$, so that the Fu-Kane index is given by 
$\nu=\nu_\Gamma\nu_{T}\nu_{X_1}\nu_{L_1}$.
Consider a Kramers pair of states at $X_1$ together with its two degenerate $\hat{C}_3$ partners at $X_2$ and $X_3$. Out of a linear combination of these states, one can construct one Kramers pair with $\hat{C}_3$ eigenvalue $-1$, and two Kramers pairs with eigenvalues $\mathrm{exp}(\pm\mathrm{i}\pi/3)$. This is shown explicitly in the Supplementary Information. When taking the Kramers pair at $X_1$ together with its degenerate partners at $X_2$ and $X_3$ to have negative inversion eigenvalue, these $\hat{C}_3$ symmetric linear combinations also have negative inversion eigenvalue. Thus, a band inversion at $X_i$ as measured by the Fu-Kane formula induces a single band inversion in the $-1$ subspace, and two (which equals no) band inversions in the $\mathrm{exp}(\pm\mathrm{i}\pi/3)$ subspace. The same holds for the $L_i$ points. We conclude that the total band inversion in the occupied subspaces with $\hat{C}_3$ eigenvalues $-1$ and $\mathrm{exp}(\pm\mathrm{i}\pi/3)$ are given by
\begin{equation}
\nu^{(\pi)}=\nu_\Gamma^{(\pi)}\nu_{T}^{(\pi)}\nu_{X_1}\nu_{L_1},
\qquad
\nu^{(\pm \pi/3)}=\nu_\Gamma^{(\pm\pi/3)}\nu_{T}^{(\pm\pi/3)},
\label{eq: topo index}
\end{equation}
 respectively.
We then distinguish three cases: 
(i) $\nu^{(\pi)}=\nu^{(\pm \pi/3)}=+1$ for a trivial insulator, 
(ii) $\nu=\nu^{(\pi)}\nu^{(\pm \pi/3)}=-1$ for a $\mathbb{Z}_2$ topological insulator, and
(iii) $\nu^{(\pi)}=\nu^{(\pm \pi/3)}=-1$ for a HOTI. 

Thus far, our considerations apply to all crystals with TRS, $\hat{C}_3$ and $\hat{I}$. We now evaluate the above topological index for elementary bismuth, crystallizing in space group $R\bar{3}m$, No. 166, which possesses these symmetries [see Fig.~\ref{fig: rod}~b)]. 
Even though bismuth is not an insulator, there exists a direct band gap separating valence bands from conduction bands [see Fig.~\ref{fig: rod}~e)]. This allows us to evaluate the indices $\nu^{(\pi)}$ and $\nu^{(\pm \pi/3)}$ for the valence bands.
We do so with the group characters obtained from first principle calculations (see methods). The result is $\nu^{(\pi)}=\nu^{(\pm \pi/3)}=-1$, which derives from $\nu^{(\pi)}_T=\nu^{(\pm \pi/3)}_T = -1$, i.e., there is a $\hat{C}_3$-graded double band inversion at the $T$ point. Hence, bismuth is a HOTI according to the topological index defined above (if we neglect the fact that it has a small electron and hole pocket).

As a second approach, we employ the formalism of elementary band representations~\cite{Bradlyn17,Vergniory17,elcoro2017double,Cano17-2,Bradlyn17-2,Cano17} (EBR) to demonstrate the nontrivial topology. Since there is always an energy separation between valence and conduction bands, we restrict our consideration to the three doubly-degenerate valence bands shown in red in Fig.~\ref{fig: rod}~e). In particular, we checked explicitly that the set of all bands at lower energy than these is topologically trivial. At TRIMs the eigenvalues of all symmetry operators have been computed (see methods). Referring to the character tables in the Bilbao Crystallographic Server (BCS)~\cite{elcoro2017double}, we  assign to all the bands their corresponding irreducible representations. The results of the eigenvalue calculations are listed in the Supplementary Information, Sec.~\ref{sec: irreps}. They show that the valence bands can not be decomposed into any linear combination of physical EBRs (pEBR, which are EBRs that respect TRS). 
It is the main result of Ref.~\onlinecite{Bradlyn17}, that if such a decomposition is not possible, the electronic band structure of bismuth has to be topological and without a description in terms of exponentially localized Wannier states, in contraposition to the conclusion drawn from Fu-Kane's parity criterion~\cite{FuKane2007}. To understand which symmetry protects this topological phase, we are repeating the symmetry eigenvalue calculation with an artificially lowered symmetry. The representative elements of point group $\bar{3}m$ are $\hat{C}_3$ around the $z$ axis (denoted 3 in the space group names), $\hat{I}$ (denoted by overbar), two-fold rotational symmetry about the $y$ axis (denoted 2), and mirror symmetry with respect to the $x$-$z$-plane (denoted $m$). After lowering the space group $R\bar{3}m$~(166) to $R3m$~(160) or $R32$~(155), a similar EBR analysis within the symmetry-reduced space groups shows that the valence bands can be decomposed into pEBRs in this case, indicating that they are topologically trivial. Therefore, neither two-fold rotation nor mirror symmetry protects the nontrivial topology of bismuth. In contrast, as long as $\hat{I}$ is preserved, lowering it to space group $R\bar{3}$~(148), the valence bands are still topological in the sense that they can not be decomposed  into pEBRs in space group 148. We conclude that the nontrivial topology is protected by $\hat{I}$ (in combination with the three-fold rotation). Notice that the rhombohedral lattice always respects the three-fold rotational symmetry. Since we learned from topological quantum chemistry that the bulk bands have no Wannier description, we expect the presence of spectral flow in Bi, and hence protected gapless modes on its boundaries. Since we know the surfaces of bismuth to be non-topological, these gapless boundaries must be hinges. This is compatible with previous works showing that Bi (111) bilayers (possibly on a substrate) host one-dimensional edge channels.~\cite{Murakami06,Takayama15}

Note that when changing the parameters of the tight binding-model of bismuth\cite{LiuAllen95} slightly, it undergoes a transition from a second-order to a first-order topological insulator\cite{Ohtsubo16}. However, we confirmed the higher-order character of bismuth that is suggested by the original tight-binding model parameters\cite{LiuAllen95} independently by performing first-principle calculations, as well an analysis in the framework of topological quantum chemistry. In particular, we took into account all occupied bands of bismuth up to its momentum-dependent energy gap. This is important since it has been shown that bands far away from this gap still contribute significantly to measurable effects, such as the unusually large $g$-factor of holes\cite{Fuseya15}.

%Having established the nontrivial bulk topology and the symmetries protecting it, we proceed by analyzing which topological boundary modes it implies.

\textit{Bulk-boundary correspondence} --- 
We present a direct calculation which let us conclude that a TRS system with $\nu^{(\pi)}=\nu^{(\pm \pi/3)}=-1$ has to have hinge modes for terminations of the crystal that globally respect inversion symmetry or further symmetries. We consider a crystal of hexagonal shape [see Fig.~\ref{fig: rod}~c)] which preserves $\hat{C}_3$ rotational and inversion symmetry. The steps outlined here in words are explicitly demonstrated using a Dirac model in the Supplementary Information, Sec.~\ref{sec: bulk-boundary}. 
We think of the insulator with $\nu^{(\pi)}=\nu^{(\pm \pi/3)}=-1$ as a superposition of two topological insulators, one in each of the independent $\hat{C}_3$ subspaces. Consider adiabatically turning off any coupling between these two subspaces, while preserving the bulk gap. The resulting system has two Dirac cones (i.e., a Dirac theory represented by $4\times 4$ matrices) on all surfaces of the crystal. Next, we seek to gap these surface Dirac cones by weakly coupling the two $\hat{C}_3$ subspaces. We want to do so while preserving the TRS, $\hat{C}_3$, and $\hat{I}$ of the crystal.
Of these, TRS is the only constraint that acts locally on a given surface. From the representation theory of the 2D Dirac equation, one finds that for a TRS that squares to $-1$, as required for spinful electrons, there exists a unique mass term $m$ that gaps the two Dirac cones in a time-reversal symmetric way. It remains to study how this mass term transforms under $\hat{C}_3$ and $\hat{I}$ to determine its relative sign between different surfaces of the crystal. Relative to the kinetic part of the surface Dirac theory, $m\to -m$ under inversion and $m\to +m$ under $\hat{C}_3$ 
(see Sec.~\ref{sec: bulk-boundary} of the Supplementary Information for details). As a result, the sign of the mass term alternates between adjacent lateral surfaces of the hexagonal crystal [see Fig.~\ref{fig: rod}~c)]. Each change of sign in the mass term is a domain wall in the Dirac theory and binds a Kramers pair of modes propagating along it. These are the one-dimensional hinge modes of the HOTI. The sign of the mass term on the top and bottom surface is not universally determined so that both patterns of hinge modes shown in Fig.~\ref{fig: rod}~c) are compatible with the bulk topology of  $\nu^{(\pi)}=\nu^{(\pm \pi/3)}=-1$ (in a real system, the particular electronic structure determines which pattern has lower energy). Apart from this ambiguity, the argument presented here solely rests on the nontrivial bulk topology and is independent of the exact form of the surface electronic structure, as long as the surface is gapped while preserving the respective symmetries. This constitutes the generalized topological bulk-boundary correspondence characteristic of a HOTI, where the existence of one-dimensional hinge modes directly follows from the 3D bulk topology. 
The HOTI's bulk-boundary correspondence requires that these hinge modes are locally stable under time-reversal symmetric perturbations that preserve the bulk and surface gaps. From this requirement, we can understand the $\mathbb{Z}_2$ topological character of the phase: the minimal TRS surface manipulation is the addition of a 2D TI to one surface of the hexagonal nanowire. This would permit hybridizing and gapping out of the pair of hinge modes adjacent to the surface. However, to comply with $\hat{I}$ and $\hat{C}_3$, the same 2D TI has to be added to every surface, thus leaving the Kramers pairs of modes intact at each hinge. We conclude that a single Kramers pair of modes at each hinge is stable under all symmetry-preserving surface perturbations. In fact, such a Kramers pair is locally stable under small perturbations even when the spatial symmetries are broken, e.g., by introducing disorder into the sample, as long a TRS is preserved. The only way to remove it is to annihilate it with another Kramers pair coming from another hinge, which cannot be achieved with just a small perturbation. The higher-order hinge modes of a 3D HOTI are therefore just as stable as the edge modes of a first-order TRS topological insulator in 2D.
We further exemplify these results with a tight-binding model, defined in Sec.~\ref{sec: TB} of the Supplementary Information, whose hinge states are shown in Fig.~\ref{fig: rod}~d),f),g). Note that our tight-binding model is topologically equivalent to a realistic model~\cite{LiuAllen95} of bismuth, however it is easier to interpret in the sense that it does not have metallic bulk and surface states that would obscure the hinge modes in the electronic structure plots we present here. It also has fewer orbitals per unit cell, which makes 3D simulations of large systems feasible.

We now turn to experimental data that support our higher-order bulk-boundary correspondence in bismuth. Even though bismuth is metallic in the bulk and on the surface, only its topological hinge states are protected against scattering by weak disorder as compared to trivial surface states, for example. We expect hinge states between (i) the top surface [which is denoted (111) in the primitive unit vectors] and three of the six lateral surfaces and (ii) between adjacent lateral surfaces.
The geometry of the samples was more amenable to study the hinge states of type (i), as we outline below.

\begin{figure}
     \centering
   \includegraphics[width=\linewidth]{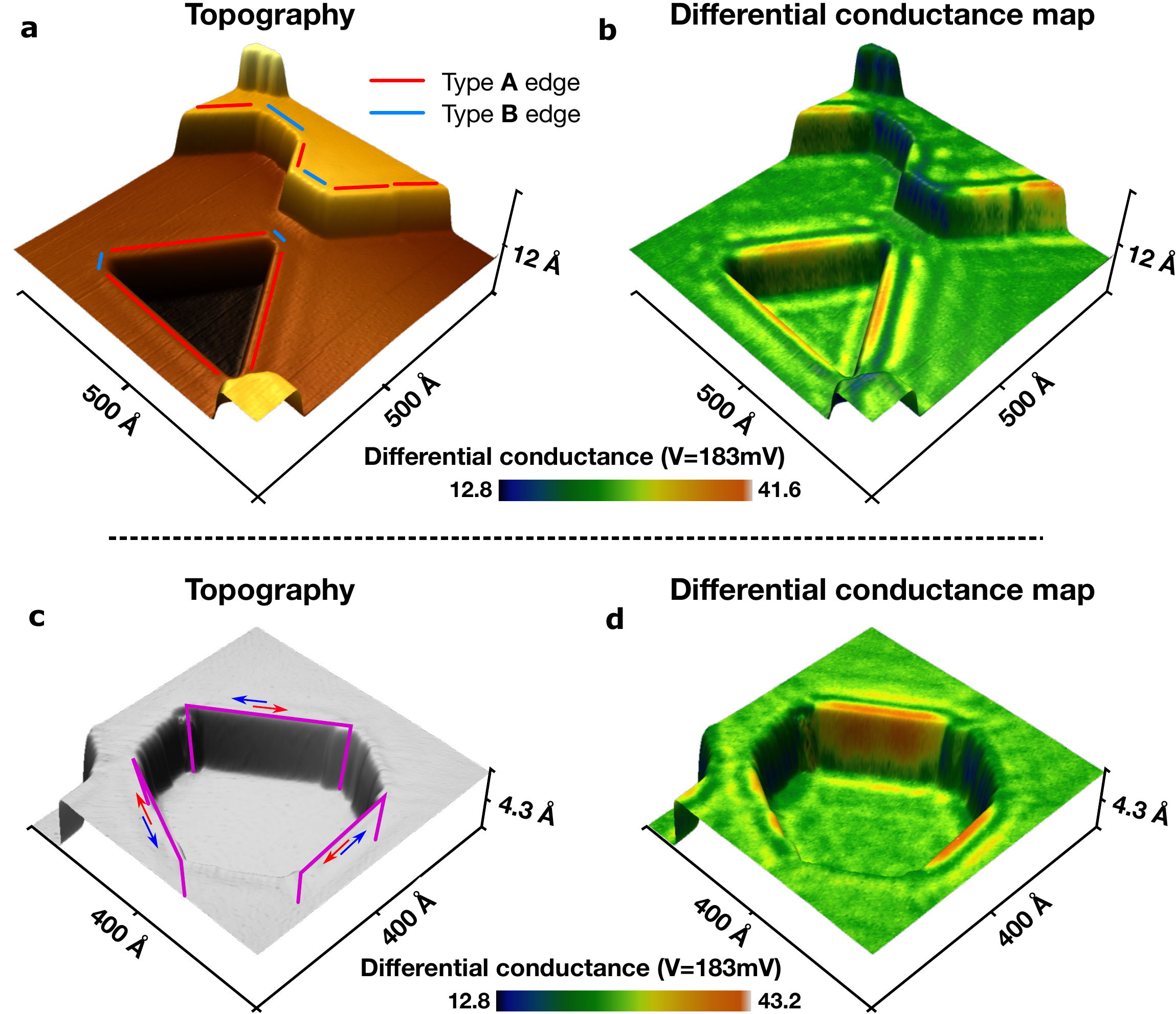}
     \caption{
Experimental observation of the alternating edge states on a bismuth (111) surface perpendicular to its trigonal axis. 
a) 3D rendered topographic image of the bismuth (111) surface. The red (type A) and blue (type B) lines then indicate the types of edge, which are along bisectrix axes. Note that the edges of type B in this particular pit geometry are much shorter than edges of type A, while still large enough to be experimentally accessible.
b) Differential conductance map at the van Hove singularity energy ($V=183$~meV) of the one-dimensional edge states. In contrast to the type B edges, all the type~A edges exhibit localized high conductance. 
c) Topographic image of a hexagonal pit on a bismuth (111) surface. The hinge modes are schematically shown as purple lines. Blue and red arrows indicate the flow of the spin-momentum locked hinge modes. 
d) Differential conductance map simultaneously acquired with the topographic data from c), showing high conductance at every other edge of the hexagonal pit. 
		}
   
      \label{BiSTM}
     \end{figure}

\textit{STM experiment} --- 
With a STM, we studied the electronic structure of step edges on the (111) surface of bismuth.  Due to the buckled honeycomb structure of the bismuth bilayer along the [111] trigonal direction, STM topographic images of the (111) plane of bismuth show bilayer steps with two different types of bisectrix edges: type A and type B [marked as red and blue lines in Fig.~\ref{BiSTM}~a)]. We highlight two structures of triangular and nearly hexagonal shape [Fig.~\ref{BiSTM}~a) and~c)].
In particular the step edge in Fig.~\ref{BiSTM}~c) can be seen as (the negative of) a one bilayer tall version of the crystals shapes shown in Fig.~\ref{fig: rod}~c). We thus expect hinge states at either the type A or the type B edges due to the higher-order topology. (All A type and all B type edges are mutually equivalent due to the $\hat{C}_3$ rotational symmetry of the bismuth (111) surface.) Indeed, we observe strongly localized edge states only at type A edges in Fig.~\ref{BiSTM}~b) and ~d), which display the differential conductance map overlaid on top of the topographic data to illuminate the edge states at the van Hove singularity energy of the bismuth edge states. A previous experimental study~\cite{Drozdov2014} showed a one-dimensional van Hove singularity of the edge states ($E=183$~meV) and quasi-particle interference of the spin-orbit locked edge states. The same study demonstrated the absence of $k$ to $-k$ scattering for these states. These experimental observations and model calculations strongly suggest that the edge states are living in the momentum dependent energy gap of the bismuth (111) surface states~\cite{Drozdov2014}. Every other edge of a hexagonal pit exhibits localized edge states and these edge states are discontinued at the corner where type A and type B edge meet [Fig.~\ref{BiSTM}~c) and~d)]. This feature remarkably reproduces the hinge modes calculated for the hexagonal nanowire as shown in Fig.~\ref{fig: rod}~d).

\begin{figure}
     \centering
   \includegraphics[width=\linewidth]{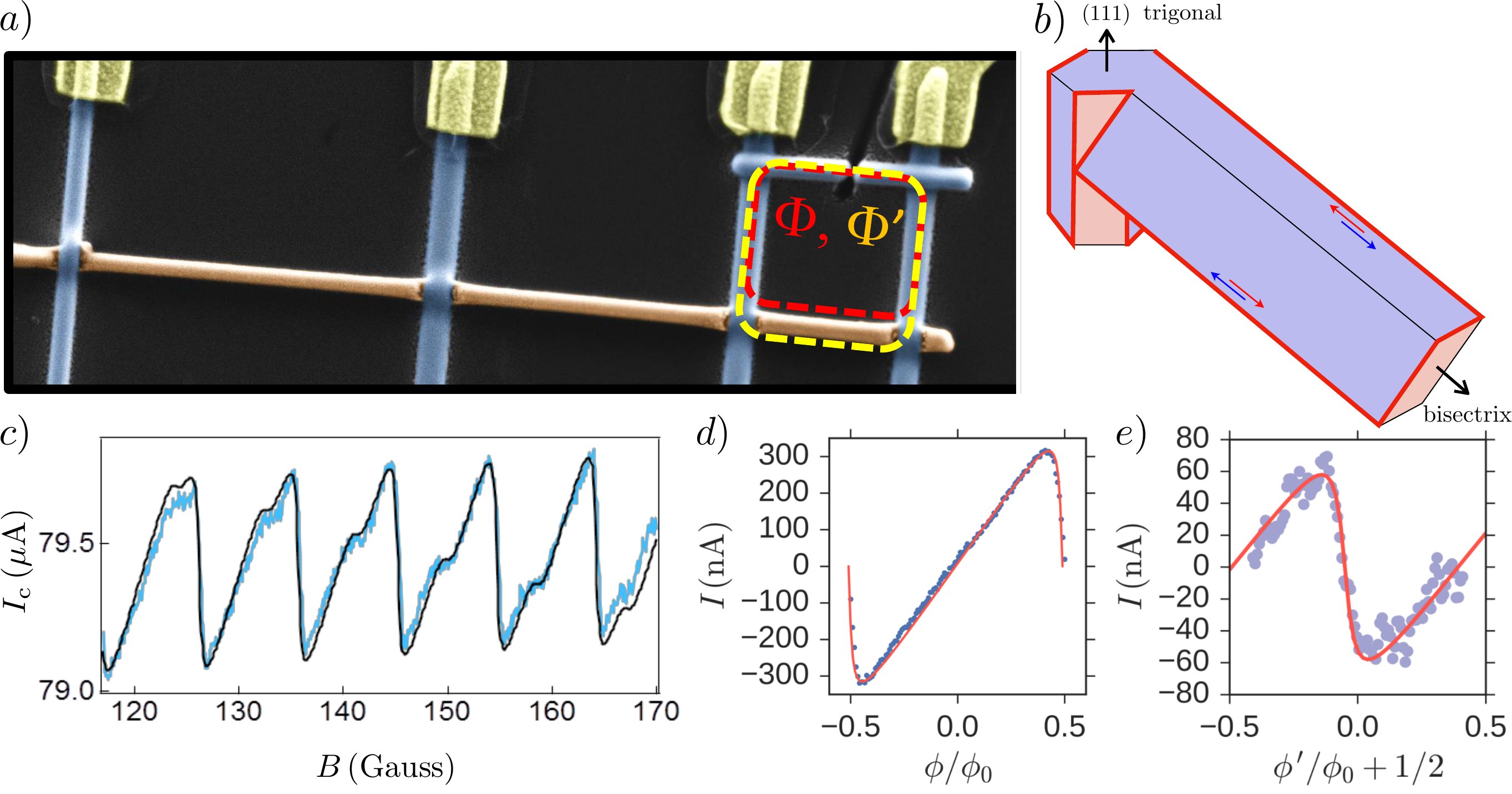}
     \caption{
     Evidence for hinge states from Josephson-interference experiments.
     a) Single-crystal bismuth nanowire (colored in brown) connected to superconducting electrodes (colored in blue). The wire has a parallelogrammatic cross-section. Its orientation along one of the bisectrix axes of bismuth was determined by electron diffraction, showing evidence of (111) facets parallel to the substrate. The $\SI{1.4}{\micro\meter}$ long, rightmost section of the wire, in parallel with a superconducting weak link, forms an asymmetric SQUID. 
     b)
		Schematic representation of the investigated bismuth nanowire of parallelogrammatic cross-section described above, indicating (red lines) the position of the experimentally  identified topological hinge states in relation to the hinge states determined theoretically in a bismuth sample of hexagonal symmetry oriented along the trigonal [111] axis.
     c)
     The magnetic field dependence of the critical current shown  is modulated by the current phase relation of the bismuth Josephson junction (whose critical current is much lower than the superconducting weak link).
		This current phase relation can be decomposed into the sum of two sawtooth waves d) and e) of different periods corresponding respectively to the internal and external area of the SQUID $\Phi$ and $\Phi'$ shown in a).
		}
      \label{BiSQUID}
     \end{figure}

\textit{Transport experiment} --- 
We exploited proximity-induced superconductivity to reveal ballistic hinge states along monocrystalline bismuth nanowires \cite{Li2014-2,Murani17}. 
 When these (non superconducting) nanowires are connected to superconducting contacts (implementing a superconductor/bismuth nanowire/superconductor  or S/Bi/S Josephson junction),  a supercurrent runs through them at low temperature. Our experiments unambiguously demonstrate that %found unexpected features, that indicate that
 the supercurrent flows via extremely few narrow one-dimensional channels, rather than via the entire surface or bulk of the nanowire. 
The experimental indications are the following: i) Periodic oscillations of the critical current through the nanowires caused by a magnetic field, with a period corresponding to one magnetic flux quantum through the wire section perpendicular to the field \cite{Li2014-2,Murani17}. Such oscillations indicate interference between two supercurrent-carrying paths located at the nanowire edges\cite{Hart14} (see also the Supplementary Material), since a uniform current density in such a long narrow wire would produce instead a monotonously decaying critical current. ii) The supercurrent flowing through the nanowire persists to extremely high magnetic fields, up to several Teslas in some samples. Since the orbital dephasing due to a magnetic flux through the supercurrent-carrying channel area destroys the induced supercurrent, this indicates that the channels are extremely narrow spatially.
%The supercurrent resists to very high magnetic field limited by the critical field of the S electrodes. Moreover  in  samples  (most of them of uncontrolled orientation) periodic oscillations could be detected with a period  related to the area of the sample perpendicular to magnetic field \cite{Li2014-2,Murani2017}. 
iii) Finally, we have recently provided a direct signature of ballistic transport along those one-dimensional channels, by measuring the supercurrent-versus-phase relation (also called current phase relation, or CPR) of  the S/Bi/S junction.  This was done by inserting the bismuth nanowires into  an asymmetric superconducting quantum interference device (SQUID) configuration\cite{DellaRocca07,Murani17}. %We have exploited the extreme sensitivity of the relation between the supercurrent through a nanostructure and the superconducting phase difference at its ends. 
Whereas tunneling or diffusive transport give rise to the usual nearly sinusoidal current phase relation of superconductor/normal metal/superconductor Josephson junctions, the sharp sawtooth-shaped current phase relation we found instead, demonstrates that transport occurs ballistically along the wire. The scattering probability $p$ was estimated to be $0.1$ along the $\SI{1}{\micro\meter}$ long bismuth wire from
the harmonics content of this current phase relation (where the $n$th harmonic decays like $(1-p)^{2n}/n$). This leads to a lower bound of the mean free path $l_e$ along these edges equal to $\SI{10}{\micro\meter}$, much longer than the value $l_e= \SI{0.1}{\micro\meter}$ determined for the surface states. This surprising  result is explained by the dominant contribution of the topologically protected hinge states to the supercurrent. Indeed, the supercurrent carried by a diffusive channel is $(L/l_e)^2 \approx 100$ times smaller than the supercurrent carried by a ballistic channel ($l_e$ and $L$ are the elastic mean free path and wire length, respectively).  The position of the edge states can be deduced from the periodicity of the SQUID oscillations, which is inversely proportional to the area enclosing the flux. In a sample of parallelogrammatic cross-section whose geometry and orientation was precisely determined, we detected a beating of two paths enclosing different fluxes $\Phi$ and $\Phi'$ [see Fig.~\ref{BiSQUID}~a)].\cite{Murani17} This demonstrated that the edge states are located along the two acute edges of the (111) facets. Those edges coincide with the expected hinge states perpendicular to the trigonal [111] axis [see Fig.~\ref{BiSQUID}~b)].
The contribution of each path  was extracted and is shown in Fig.~\ref{BiSQUID}~d) and~e). The supercurrents carried by the two hinges differ by a factor of four. This can be explained by a difference in the quality of the contact to these hinge states: The top hinges of the wire have been more severely etched than the bottom ones during the deposition of the superconducting electrodes [see Fig.~\ref{BiSQUID}~a)]. This strong etching reduces the coupling of edge states to the superconducting contacts and the supercurrent is decreased even though the ballistic nature is unaffected.

Comparing Fig.~\ref{BiSQUID}~d) and Fig.~\ref{fig: rod}~c), we note that one of the two hinges on top of the nanowire must be of A type and the other one of B type (the same is true for the bottom two hinges). Our observation of a ballistic channel at one of these hinges at the top, and one at the bottom of the nanowire, is thus in line with the theoretical expectation from the higher-order topology of bismuth.

\textit{Summary ---}
The bismuth-antimony alloy, Bi$_{1-x}$Sb$_x$, was the first material realization of a 3D TI~\cite{FuKane2007,Hsieh09}. 
The composition $x$ was used to interpolate between the bismuth without band inversion  
and the band inverted antimony. 
In this work, we demonstrated theoretically that the allegedly trivial end of this interpolation, bismuth,  
has in fact a 3D topological band structure as well. It is a HOTI with helical hinge states. We presented two complementary pieces of experimental evidence supporting this result, using STM and Josephson-interferometry measurements. 
The type of hinge states discussed here may be used for lossless electronic transport due to their local protection from backscattering by TRS disorder. Further applications include spintronics, due to their spin-momentum locking, and -- when proximitized with superconductivity --  topological quantum computation. For the latter, a nanowire with hexagonal cross-section may provide a particularly convenient way of building a hexon -- a group of six Majorana states, one at each hinge. Hexons have been proposed as building blocks for a measurement-only quantum computer~\cite{StationQHexons}.

\textit{Data Availability ---}
The data that support the plots in Figs.~\ref{fig: rod},\ref{BiSTM},\ref{BiSQUID} within this paper and other findings of this study are available from the corresponding author upon reasonable request. The information on elementary band representations is available on the Bilbao crystallographic server\cite{Aroyo2011183}.

%\vspace{1cm}%beautifier
%%%%%%%%%%%%%%%%%%%%%%%%%%%%%%%%%%%%%%%%%%%%%%%%%%%%
\textit{Acknowledgments ---}
FS and TN acknowledge support from the Swiss National Science Foundation (grant number: 200021\_169061) and from the European Union's Horizon 2020 research and innovation program (ERC-StG-Neupert-757867-PARATOP). MGV was supported by IS2016-75862-P national project of the Spanish MINECO. AMC wishes to thank the Kavli Institute for Theoretical Physics, which is supported by the National Science Foundation under Grant No. NSF PHY-1125915, for hosting during some stages of this work. AM, SS, AYK, RD, HB, and SG thank Manuel Houzet who drew their attention to recently published work on higher-order topological insulators. They were supported by the ANR grants DIRACFORMAG, MAGMA and JETS. 
AY acknowledges support from NSF-MRSEC programs through the Princeton Center for Complex Materials DMR-142054, NSF-DMR-1608848, and ARO-MURI program W911NF-12-1-046.
BAB acknowledges support for the analytic work Department of Energy de-sc0016239, Simons Investigator Award, the Packard Foundation, and the Schmidt Fund for Innovative Research. The computational part of the Princeton work was performed under NSF EAGER grant DMR -- 1643312, ONR - N00014-14-1-0330, ARO MURI W911NF-12-1-0461, NSF-MRSEC DMR-1420541. 

\textit{Author contributions ---}
F.S., A.M.C., B.A.B, and T.N. carried out the theoretical analysis and model calculations. Z.W. and M.V. performed the first-principle calculations and topological quantum chemistry analysis. A.M., S.S., A.Y.K., R.D., H.B., and S.G. conceived and carried out the transport experiments including crystal growth. 
S.J., I.D., and A.Y. conceived and carried out the STM/STS experiments.

\section*{Methods}

\textit{First-principle calculations} ---
We employed density functional theory (DFT) as implemented in the Vienna Ab Initio Simulation Package (VASP)\cite{vasp1,vasp2,vasp3,vasp4}. The exchange correlation term is described according to the Perdew-Burke-Ernzerhof (PBE) prescription together with
projected augmented-wave pseudopotentials\cite{PBE-1996, PBE} and the spin-orbit interaction included. For the self-consistent calculations we used a $12 \times 12 \times 12$ $\bs{k}$-points mesh for the bulk band structure calculations. The eigenvalues of the symmetry transformations were deduced from the matrix representations of the respective symmetry operation calculated using the Bloch eigenstates from VASP.

\textit{STM experiment} ---
Bismuth crystals were cleaved at room temperature in ultra-high vacuum conditions and the cleaved samples were cooled down to a temperature of 4~K at which scanning tunneling microscopy (STM) and spectroscopy (STS) measurements were carried out.
The cleaved bismuth crystal exhibits a (111) plane of the bismuth rhombohedral structure [which is the (001) plane of the bismuth hexagonal structure]. The topographic data and the differential conductance maps were. 
 For STM measurements, a mechanically sharpened platinum-iridium tip was used, and electronic properties of the probe tip were characterized before the experiments on bismuth by checking a reference sample. Differential conductance maps [Fig.~\ref{BiSTM}~b) and~d)] are taken simultaneously with topographic data at the van Hove singularity energy ($V=183$~meV) of the bismuth edge states using a lock-in amplifier with an oscillation of 3~meV and with $I=3.5$~nA. The data shown in this manuscript is reproduced on many step edges of Bi (111) with atomically different tips. All of the islands on the Bi (111) surface show the expected step height of 4~\AA~for bismuth bilayers and all of the extended edges are identified as zigzag structures of either A type or B type. A type and B type edges are equivalent in the hexagonal nanowire geometry as described in the main manuscript [Fig.~\ref{fig: rod}~c)], however, the existence of the Bi (111) surface under the bismuth bilayer breaks the inversion symmetry, and A as well as B type edges can be identified in STM measurements. Only A type edges show the spectroscopic feature of a sharp peak at 183~meV which is the van Hove singularity energy of the one-dimensional edge state. Quasi-particle interference (QPI) measurements reveal that this edge state is continuously dispersing down to the Fermi level and starts to merge with the surface states at the momentum where the surface gap closes~\cite{Drozdov2014}. This spectroscopic feature of geometric confinement only at A type edges resembles the topological hinge modes expected for the hexagonal nanowire, as discussed in the main text.

\textit{Transport experiment} ---
The nanowires grew during slow sputtering deposition of high purity bismuth on a slightly heated silicon substrate. High resolution transmission electron microscopy (TEM) indicates high quality single crystals, of hexagonal or rhombohedral cross-sections, with clear facets. The facet widths are typically 50 to 300~nm wide. Resistance measurements show that transport in the normal state (i.e., when contacts to the nanowires are not superconducting) occurs predominately due to surface states,  with an elastic mean free path of the order of 100~nm.

\bibliography{Ref-Lib}

%\cleardoublepage
\clearpage
\beginsupplement
\begin{widetext}
\section*{Supplementary Information: Higher-Order Topology in Bismuth}

\subsection{Bulk-boundary correspondence}
\label{sec: bulk-boundary}
In this section, we derive the bulk-surface-hinge correspondence for a HOTI with TRS, $\hat{C}_3$ rotation, and $\hat{I}$.
To that end, we consider a Dirac model representation of a HOTI based on the Bernevig-Hughes-Zhang (BHZ) model for 3D topological insulators~\cite{Bernevig06}. Our model corresponds to a continuum limit of the tight-binding model considered in the next section (when expanded around the $T$ point in the Brillouin zone) and shares all relevant topological features with elementary bismuth.
It has eight bands and is written in the orbital basis
$\{
|p_-\uparrow\rangle, 
|d_-\downarrow\rangle,
|p_+\downarrow\rangle, 
|d_+\uparrow\rangle,
|p_+\uparrow\rangle, 
|d_+\downarrow\rangle,
|p_-\downarrow\rangle, 
|d_-\uparrow\rangle
\}$, where $p_\pm=p_x\pm\mathrm{i}p_y$ and $d_\pm=d_{xy}\pm\mathrm{i}d_{x^2-y^2}$. 
We are interested in preserving $\hat{C}_3$ rotation symmetry around the $z$ axis (our spin quantization axis). It is represented as
\begin{equation}
\label{eq: c3symmetryrep}
C^z_3
%=\mathrm{diag}
%\left(
%e^{\mathrm{i}\pi/3},e^{-\mathrm{i}\pi/3},e^{-\mathrm{i}\pi/3},e^{\mathrm{i}\pi/3},
%-1,-1,-1,-1
%\right).
=e^{-\mathrm{i} \frac{2\pi}{3} s}, \quad
s = \mathrm{diag}\left(\frac{1}{2}, \frac{5}{2}, -\frac{1}{2}, -\frac{5}{2}, \frac{3}{2}, \frac{3}{2}, -\frac{3}{2}, -\frac{3}{2} \right).
\end{equation}
Inversion and TRS are represented by 
$I=\sigma_0\otimes\sigma_0 \otimes \sigma_3$ (since $p$ and $d$ orbitals have opposite inversion eigenvalues) and
$T=\sigma_0\otimes\sigma_2 \otimes \sigma_0 \mathit{K}$, where $\mathit{K}$ denotes complex conjugation, $\sigma_0$ the $2\times 2$ identity matrix, and $\sigma_1, \sigma_2, \sigma_3$ the three Pauli matrices. Here, the tensor product is defined such that $\textrm{diag} \left(\sigma_3 \otimes \sigma_0 \otimes \sigma_0 \right) = (1, 1, 1, 1, -1, -1, -1, -1)$.

The Hamiltonian for two $\hat{C}_3$ eigenspaces which are (as of yet) not connected is given by
 \begin{equation}
H(\bs{k}) = H_1(\bs{k}) \oplus H_3(\bs{k}),
\label{eq: continuum H}
\end{equation}
where $\bs{k}$ is measured from the $T$ point in the BZ, and the blocks are defined as
\begin{equation}
H_j(\bs{k}) =
\begin{pmatrix} M(\bs{k}) & A_j k_+^j & 0 & \tilde{A} k_z \\ 
A_j k_-^j & -M(\bs{k}) & \tilde{A} k_z & 0 \\
0 & \tilde{A} k_z & M(\bs{k}) & - A_j k_-^j \\
\tilde{A} k_z & 0 & - A_j k_+^j & - M(\bs{k})
\end{pmatrix},
\end{equation}
for $j=1,3$.
Here $k_\pm=k_x\pm\mathrm{i}k_y$ and $M(\bs{k})=M_0-M_1 \bs{k}^2$.  $M_0$, $M_1$ and $\tilde{A}$ are free parameters, which we choose to be equal in both $\hat{C}_3$ eigenspaces for simplicity. For the same reason, we only consider the case $A_1 = A_3 \equiv A$ in this section. We also make the choice $\mathrm{diag}(1,-1,1,-1)$ for the matrix multiplying the function $M(\bs{k})$ without loss of generality. In $H_1(\bs{k})$ we recognize the BHZ model. To motivate this ansatz, consider for example the matrix element
\begin{equation}
\bra{p_-\uparrow} H (\bs{k}) \ket{d_-\downarrow} = \bra{p_-\uparrow} (C^z_3)^{-1} H (C^z_3 \bs{k}) C^z_3 \ket{d_-\downarrow} = e^{-\mathrm{i} \frac{4\pi}{3}} \bra{p_-\uparrow} H (C^z_3 \bs{k}) \ket{d_-\downarrow},
\end{equation}
where $\hat{C}_3$ symmetry requires that (to lowest order) it has to be proportional to either $k_+$, which satisfies $C_3 k_+ = e^{-\mathrm{i} \frac{2\pi}{3}} k_+$, or $k_-^2$, which satisfies $C_3 k_-^2 = e^{\mathrm{i} \frac{4\pi}{3}} k_-^2$, i.e., $\hat{C}_3^z$-symmetry requires this matrix element to be proportional to $k_+^n$ with $n=1 \mod 3$. Since $p$ $(d)$ orbitals are odd (even) under inversion, we are restricted to odd powers of $k$ by inversion symmetry, therefore only $k_+$ is admissible.
Likewise, this off-diagonal coupling has to be modified to $k_+^3$ in $H_3(\bs{k})$ due to inversion symmetry and the specific representation of $\hat{C}_3$ in Eq.~\eqref{eq: c3symmetryrep}.

Each Hamiltonian block has a band inversion at $\bs{k}=0$ and thus the full model represents a HOTI as we showed in the main text. In the case of bismuth, were the band inversion happens at the $T$ point, $\bs{k}$ should be understood as the momentum measured from that point. When discussing the bulk-boundary correspondence we will introduce couplings between the two blocks $H_1(\bs{k})$ and $H_3(\bs{k})$.

To analytically solve for the hinge modes, we consider Hamiltonian~\eqref{eq: continuum H} on a cylinder of radius $r_0$ with the cylinder axis parallel to $z$. Each of the independent blocks of Hamiltonian~\eqref{eq: continuum H} is a 3D TI and should thus support one Dirac cone on the surface. We will first solve for these Dirac states and then gap them out via mutual coupling. We will show that for this coupling term to comply with the symmetry requirements it has to have zeros along six lines on the cylinder surface that run parallel to $z$ and correspond to the hinge states. Note that while the geometry that we chose is a cylinder with continuous rotational symmetry, the Hamiltonian~\eqref{eq: continuum H} has only $\hat{C}_6$ rotational symmetry.
%To solve for surface states, we choose periodic boundary conditions in $z$-direction.
First, we only solve for the states that form the degeneracy point of the surface Dirac electrons, i.e., those with vanishing transversal momentum on the surface. We thus set $k_z=0$ and replace $k_+ \to - \mathrm{i} e^{\mathrm{i}\phi} \partial_\rho$, where $r\equiv r_0+\rho$ is the radial coordinate of the cylinder, and $\phi$ the angular coordinate. A domain wall between the topologically nontrivial sample and the vacuum is realized by setting $M_0 \rightarrow M_0(\rho) = \bar{M} \, \mathrm{sgn}{\rho}$. We can set $M_1=0$ for the purpose of solving for surface states. The Hamiltonian $H = H_1 \oplus H_3$ is then composed of
\begin{equation}
\begin{aligned}
H_j &= 
\begin{pmatrix}
h_j & 0 \\ 0 & h_j^*
\end{pmatrix}, \\
h_j &= \begin{pmatrix} \bar{M} \, \mathrm{sgn}{\rho} & A (-\mathrm{i})^j e^{\mathrm{i} j \phi} \partial^j_\rho \\
A (-\mathrm{i})^j e^{-\mathrm{i} j \phi} \partial^j_\rho & - \bar{M} \, \mathrm{sgn}{\rho} \end{pmatrix}.
\end{aligned}
\end{equation}
Each $H_j$ has two normalizable zero-energy solutions,
\begin{equation}
\ket{v_j} = a_j
\begin{pmatrix}
1 \\ \mathrm{i}^j e^{-\mathrm{i} j \phi} \mathrm{sgn}\, \frac{\bar{M}}{A} \\ 0 \\ 0
\end{pmatrix},
\ket{w_j} = a_j
\begin{pmatrix}
0 \\ 0 \\ 1 \\ (-\mathrm{i})^j e^{\mathrm{i} j \phi} \mathrm{sgn}\, \frac{\bar{M}}{A}
\end{pmatrix},
\end{equation}
with $a_j\equiv e^{- \left|\frac{\bar{M}}{A}\right|^\frac{1}{j} |\rho|}$.
In combination, $H$ has four normalizable zero-energy solutions. When reintroducing $k_z$ and the surface (angular) momentum conjugate to $\phi$, these zero-energy states disperse as a pair of surface Dirac cones, as is expected from having a strong topological insulator in each of the two $\hat{C}_3$ subspaces. However, we can gap out these two Dirac cones by introducing a coupling between them without breaking TRS. Requiring in addition $\hat{C}_3$ and $\hat{I}$ to be preserved then imposes constraints on the real space dependence of a possible mass term $M_\mathrm{s}(\phi)$ on the surface of the cylinder. 

Let us first focus on the local constraints on such a mass term.
In order to gap out the surface Dirac cones, $M_\mathrm{s}$ should anticommute with all the surface kinetic terms that multiply the two transverse momenta on the surface. 
Further, $M_\mathrm{s}$ should couple different $\hat{C}_3$ subspaces, i.e., it must be proportional to $\sigma_i\otimes \sigma_\mu\otimes \sigma_\nu$ with $i=1,2$ when written in the original basis of Hamiltonian~\eqref{eq: continuum H}. Any other choice of $i = 0,3$ would not couple the two distinct time-reversal symmetric $\hat{C}_3$ subspaces. $M_\mathrm{s}$ should also be TRS.
Finally, $M_\mathrm{s}$ should anticommute with the bulk mass term involving $M$ in Eq.~\eqref{eq: continuum H}, which is proportional to $\sigma_0\otimes \sigma_0\otimes \sigma_3$. Otherwise it represents a competing mass and results in a gapless region near the surface.

Taking all the local restrictions into account, we find that the only allowed surface mass term is of the form 
\begin{equation}
M_\mathrm{s} (\phi) = m(\phi) \sigma_2\otimes \sigma_0\otimes \sigma_2 + \tilde{m}(\phi) \sigma_2\otimes \sigma_3\otimes \sigma_1,
\label{eq: Ms ansatz}
\end{equation}
where we discarded a possible $\rho$-dependence that is unaffected by the above restrictions. We now impose $\hat{C}_3$ symmetry. It transforms the matrices in Eq.~\eqref{eq: Ms ansatz} as
\begin{equation}
\begin{aligned}
C^z_3 (\sigma_2\otimes \sigma_0\otimes \sigma_2) \left(C^z_3\right)^{-1}
 = 
\cos \frac{2 \pi}{3} \sigma_2\otimes \sigma_0\otimes \sigma_2 - \sin \frac{2 \pi}{3} \sigma_2\otimes \sigma_3\otimes \sigma_1,\\
C^z_3 (\sigma_2\otimes \sigma_3\otimes \sigma_1) \left(C^z_3\right)^{-1}
 = 
\cos \frac{2 \pi}{3} \sigma_2\otimes \sigma_3\otimes \sigma_1 + \sin \frac{2 \pi}{3} \sigma_2\otimes \sigma_0\otimes \sigma_2.
\end{aligned}
\end{equation}
In addition, both matrices in Eq.~\eqref{eq: Ms ansatz} anticommute with inversion. To maintain $\hat{C}_3$ symmetry, we require
\begin{equation}
C^z_3 M_\mathrm{s} (\phi) (C^z_3)^{-1} = M_\mathrm{s} \left(\phi + \frac{2 \pi}{3}\right).
\end{equation}
This means that to lowest order the allowed harmonics of the coefficients $m(\phi)$ and $\tilde{m}(\phi)$ in $\phi$ that are compatible with $C^z_3$ and $I$ symmetry are given by
\begin{equation}
\begin{split}
m(\phi) =&\, m'' \sin(\phi) + m' \cos (\phi),
\\
 \tilde{m}(\phi) =&\, -m' \sin(\phi) + m'' \cos (\phi),
 \end{split}
\end{equation}
with arbitrary real coefficients $m'$ and $m''$.
Without loss of generality we may choose $m''=0$ (which amounts to fixing an origin for $\phi$ so that $\phi = 0$ corresponds to the $x$-direction). We can now project the corresponding mass term in the basis of surface Dirac states,
\begin{equation}
\begin{split}
\braket{v_1 | M_\mathrm{s}(\phi) | v_3} &\,\propto  -e^{-\mathrm{i} 4 \phi} (e^{\mathrm{i} 6 \phi} - 1) , \\
\braket{w_1 | M_\mathrm{s}(\phi) | w_3} &\,\propto   e^{-\mathrm{i} 2 \phi} (e^{\mathrm{i} 6 \phi} - 1),
\end{split}
\end{equation}
with all matrix elements not related to these by hermiticity vanishing. 
Importantly, the mass projected into the surface states vanishes at exactly six equally spaced angles due to the prefactor $(e^{\mathrm{i} 6 \phi} - 1)$. Expanding around, for instance, $\phi=0$, we have $\braket{v_1 | M_\mathrm{s}(\phi) | v_3}=\braket{w_1 | M_\mathrm{s}(\phi) | w_3} \sim  \phi$. Thus, the surface mass has a domain wall located at $\phi=0$ at which it changes sign. By $\hat{C}_3$ together with $\hat{I}$ the same is true at all $\phi$ of the form $\phi = \frac{2\pi}{3} n$, $n=0, \cdots, 5$. Since a domain wall in the mass of a pair of 2D TRS Dirac fermions binds a Kramers pair of gapless modes, the model represented by $H(\bs{k})$ has a helical pair of gapless modes at each of the hinges in a hexagonal real-space geometry that preserves $\hat{C}_3$ and $\hat{I}$, while the surfaces are gapped. This constitutes the second-order bulk-boundary correspondence of the HOTI.

\begin{figure}
\centering
\includegraphics[width=0.5\textwidth]{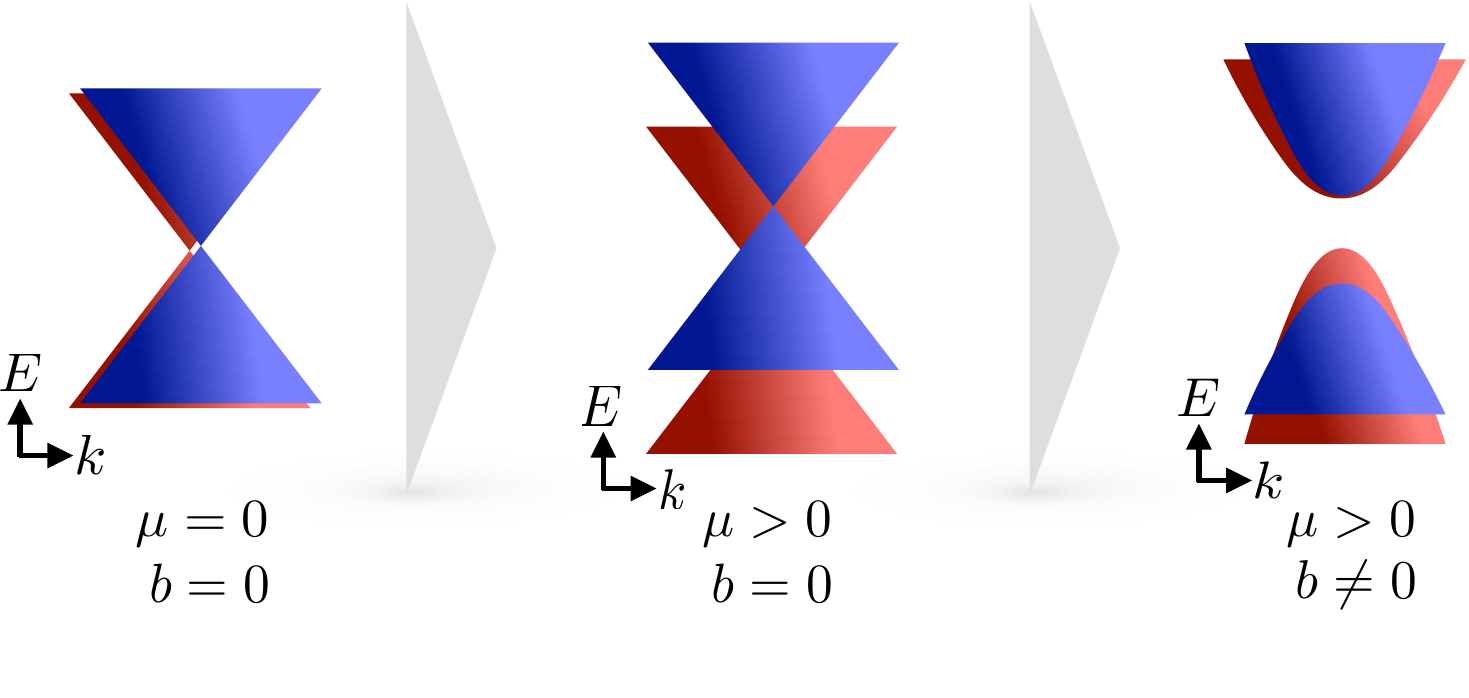}
\caption{
Effect of the surface mass term in Eq.~\eqref{eq: zDirectionMass} on the two Dirac cones on the $(001)$ surface of the continuum model given by Eq.~\eqref{eq: continuum H}. Since initially both cones are centered at the projection of the bulk BZ $T$ point onto the surface BZ, which is invariant under $\hat{C}_3$ symmetry, they cannot be directly gapped out due to their different $\hat{C}_3$ eigenvalues. However, after a relative shift in energy effected by a finite $\mu$ in Eq.~\eqref{eq: zDirectionMass}, we can gap out the resulting nodal line.
}
\label{fig: SI-Dirac-Bandstruct}
\end{figure}

Instead of considering a mass in real space, we can also gap out the surface Dirac cones due to the Hamiltonian in Eq.~\eqref{eq: continuum H} by adding a momentum-dependent mass term to the bulk theory. The corresponding bulk Hamiltonian reads
\begin{equation}
H_1(\bs{k}) \oplus H_3(\bs{k}) + \begin{pmatrix} 0 & H_{M_\rho}(\bs{k}) \\ H_{M_\rho}(\bs{k})^\dagger & 0 \end{pmatrix}, \quad
H_{M_\rho}(\bs{k}) = \begin{pmatrix} 
- \bar{a} k_+^2 & 0 & 0 & 0 \\
0 & -a k_-^2 & 0 & 0 \\
0 & 0 & a k_-^2 & 0 \\
0 & 0 & 0 & \bar{a} k_+^2
\end{pmatrix},
\end{equation}
where $a$ is a complex number that, when non-vanishing, gaps out the surface Dirac cones in the cylindrical geometry described above almost everyhwere, leaving behind six propagating hinge modes compatible with $\hat{C}_3$ and $\hat{I}$ symmetry.

Finally, we discuss the case of open boundary conditions in the $z$-direction. In several other HOTIs protected by a rotational symmetry, surfaces with normal parallel to the rotation axis are gapless~\cite{SchindlerHOTI, Fang17}. In contrast, for the case at hand the $z$-surface can be gapped. A subtlety arises from the fact that the two Dirac cones on this surface cannot be gapped by a straightforward Dirac mass term, because of a mismatch in rotation eigenvalue in the two subspaces with $\hat{C}_3$ eigenvalues $-1$ and $\mathrm{exp}(\pm\mathrm{i}\pi/3)$. However, a combination of a relative energy shift between the two subspaces and a mass term that vanishes at the Dirac node can induce a gap~\cite{Iadecola14} (see Fig.~\ref{fig: SI-Dirac-Bandstruct}). The correspondingly perturbed Hamiltonian reads
\begin{equation}
\label{eq: zDirectionMass}
H_1(\bs{k}) \oplus H_3(\bs{k}) + \begin{pmatrix} \mu\, \sigma_0\otimes\sigma_0& H_{M_z}(\bs{k}) \\ H_{M_z}(\bs{k})^\dagger & -\mu\, \sigma_0\otimes\sigma_0 \end{pmatrix}, \quad
H_{M_z}(\bs{k}) = \begin{pmatrix} 
0 & 0 & 0 & -\mathrm{i} b k_- \\
0 & 0 & \mathrm{i} \bar{b} k_+ & 0 \\
0 & -\mathrm{i} \bar{b} k_+ & 0 & 0 \\
\mathrm{i} b k_- & 0 & 0 & 0
\end{pmatrix},
\end{equation}
where $b$ is another complex number. Therefore, the full HOTI bulk Dirac theory, where all surfaces are gapped in the open geometry of a hexagonal prism, and only hinge modes remain, reads
\begin{equation}
\label{eq: SIgappedDiracEquation}
H_\mathrm{D} (\bs{k}) =H_1(\bs{k}) \oplus H_3(\bs{k}) + \begin{pmatrix} \mu & H_{M_\rho}(\bs{k})+H_{M_z}(\bs{k}) \\ H_{M_\rho}(\bs{k})^\dagger+H_{M_z}(\bs{k})^\dagger & -\mu \end{pmatrix}.
\end{equation}

\subsection{Tight-binding model}
\label{sec: TB}

In addition to the Dirac model, we also provide a tight-binding model for a HOTI. It is defined on the simple hexagonal lattice spanned by the lattice vectors $\bs{a}_1 = (1,0,0)$, $\bs{a}_2 = (-1/2,\sqrt{3}/2,0)$, and $\bs{a}_3 = (0,0,1)$, see Fig.~\ref{fig: tightbinding}~a) for the lattice structure. At each site of the lattice we place a $p_x$, a $p_y$, a $d_{xy}$, and a $d_{x^2-y^2}$ orbital, each of which has two spin states. This gives eight local fermionic degrees of freedom per unit cell. Note that this does not correspond to the case of bismuth, which has 16 relevant inequivalent local fermionic degrees of freedom per primitive unit cell and 48 degrees of freedom in the conventional unit cell~\cite{LiuAllen95}. However, our tight-binding model has the very same bulk topology as bismuth, which we obtained from DFT calculations and topological quantum chemistry, and which furthermore agrees with the bulk topology of the realistic tight-binding model\cite{LiuAllen95}. As we want to probe the topological features of bismuth, it is therefore sufficient to study the simpler model used here as long as this condition of topological equivalence is met. The model is defined via the eight-band Bloch Hamiltonian
\begin{equation}
\begin{aligned}
\label{eq: TBHamiltonian}
H_\mathrm{TB}(\bs{k}) =&\,
\begin{pmatrix}
H_\mathrm{TB, I}(\bs{k}) + \epsilon & \delta \, M_\mathrm{TB} (\bs{k}) \\ \delta \, M_\mathrm{TB} (\bs{k})^\dagger & H_\mathrm{TB, II}(\bs{k}) - \epsilon
\end{pmatrix}, \\
H_\mathrm{TB, I}(\bs{k}) =&\Gamma_1 \bigl\{m_\mathrm{I} (1 + \cos \bs{k}\cdot \bs{a}_3) - t_\mathrm{I} \left[\cos \bs{k}\cdot \bs{a}_1 + \cos \bs{k}\cdot \bs{a}_2 + \cos \bs{k}\cdot (\bs{a}_1 + \bs{a}_2) \right] \bigr\} \\&+ \lambda_\mathrm{I} \bigl[\Gamma_2 \sin \bs{k}\cdot \bs{a}_1 + \Gamma^{\mathrm{I},\mathrm{I}} _{2, 1} \sin \bs{k}\cdot \bs{a}_2 -  \Gamma^{\mathrm{I},\mathrm{I}} _{2, 2} \sin \bs{k}\cdot (\bs{a}_1 + \bs{a}_2) + \Gamma_3 \sin \bs{k}\cdot \bs{a}_3
\bigr],\\
H_\mathrm{TB, II}(\bs{k}) = &\Gamma_1 \bigl\{m_{\mathrm{II}} (1 + \cos \bs{k}\cdot \bs{a}_3) - t_{\mathrm{II}} \left[\cos \bs{k}\cdot \bs{a}_1 + \cos \bs{k}\cdot \bs{a}_2 + \cos \bs{k}\cdot (\bs{a}_1 + \bs{a}_2) \right] \bigr\}
\\&
+\lambda_{\mathrm{II}} \bigl[ \Gamma_2 \sin \bs{k}\cdot \bs{a}_1 + \Gamma^{\mathrm{II},\mathrm{II}} _{2, 1} \sin \bs{k}\cdot \bs{a}_2 - \Gamma^{\mathrm{II},\mathrm{II}} _{2, 2} \sin \bs{k}\cdot (\bs{a}_1 + \bs{a}_2) + \Gamma_3 \sin \bs{k}\cdot \bs{a}_3 \bigr]
\\ &
+ \Gamma_4 \gamma_{\mathrm{II}} \bigl[\sin \bs{k}\cdot (\bs{a}_1 + 2 \bs{a}_2) + \sin \bs{k}\cdot (\bs{a}_1 - \bs{a}_2) - \sin \bs{k}\cdot (2 \bs{a}_1 + \bs{a}_2) \bigr],
\\
M_\mathrm{TB}(\bs{k}) =&\, \Gamma_2 \bigl[\sin \bs{k} \cdot \bs{a}_1 + \sin \bs{k} \cdot (2\bs{a}_1 + \bs{a}_2)\bigr] + \Gamma^{\mathrm{I},\mathrm{II}} _{2, 1} \bigl[\sin \bs{k} \cdot \bs{a}_2 + \sin \bs{k} \cdot (\bs{a}_2 - \bs{a}_1)\bigr] 
\\&
- \Gamma^{\mathrm{I},\mathrm{II}} _{2, 2} \bigl[\sin \bs{k} \cdot (\bs{a}_1+\bs{a}_2) + \sin \bs{k} \cdot (\bs{a}_1 + 2\bs{a}_2)\bigr]
-\mathrm{i} \Gamma_5 \bigl[\cos \bs{k} \cdot \bs{a}_1 + \cos \bs{k} \cdot (2\bs{a}_1 + \bs{a}_2)\bigr] \\
&-\mathrm{i} \Gamma^{\mathrm{I},\mathrm{II}} _{5, 1} \bigl[\cos \bs{k} \cdot \bs{a}_2 + \cos \bs{k} \cdot (\bs{a}_2 - \bs{a}_1)\bigr] -\mathrm{i} \Gamma^{\mathrm{I},\mathrm{II}} _{5, 2} \bigl[\cos \bs{k} \cdot (\bs{a}_1+\bs{a}_2) + \cos \bs{k} \cdot (\bs{a}_1 + 2\bs{a}_2)\bigr],
\end{aligned}
\end{equation}
where $\Gamma_1 = \sigma_3\otimes\sigma_0$, $\Gamma_2 = \sigma_1\otimes\sigma_1$, $\Gamma_3 = \sigma_2\otimes\sigma_0$, $\Gamma_4 = \sigma_1\otimes\sigma_2$, $\Gamma_5 = \sigma_3\otimes\sigma_1$, and
\begin{equation}
\Gamma_{\mu, \nu}^{i,\ j} = \left(C^{z}_{3,i} \right) ^{\nu} \Gamma_{\mu} \left(C^z_{3,j} \right)^{-\nu}.
\end{equation}
%$\Gamma ' _{\mu} = \Gamma^{\mathrm{I},\mathrm{I}} _{\mu, 1}$
%$\Gamma * _{\mu} = \Gamma^{\mathrm{II},\mathrm{II}} _{\mu, 1}$
%$\Gamma^{ \circ} _{\mu} = \Gamma^{\mathrm{I},\mathrm{II}} _{\mu, 1}$
Here, $\mu \in \{ 1,\ \cdots , 5 \}$, $i, j \in \{\mathrm{I} ,\mathrm{II}  \}$, $\nu \in \{1,\ 2 \}$,  $C^z_{3,\mathrm{I}} = \sigma_0 \otimes e^{\mathrm{i} \frac{\pi}{3} \sigma_3}$ and $C^z_{3,\mathrm{II}} = - \sigma_0 \otimes \sigma_0$ so that the full threefold rotation symmetry is given by $C_3^z = C^z_{3,\mathrm{I}} \oplus C^z_{3,\mathrm{II}}$. Terms involving $\Gamma_1$ and $\Gamma_3$ implement intra and inter-orbital hopping, respectively, whereas the other terms correspond to various forms of spin-orbit coupling. All $\bs{k}$-dependencies involving $\bs{a}_1$ and its $C_3^z$ rotations $\bs{a}_2$, $-(\bs{a}_1 + \bs{a}_2)$ correspond to nearest-neighbor couplings, while terms involving $(\bs{a}_1 + 2 \bs{a}_2)$ and its $C_3^z$ rotations $-(2 \bs{a}_1 + \bs{a}_2)$, $(\bs{a}_1 - \bs{a}_2)$ correspond to next-nearest neighbor couplings. To enhance readability we have simplified the expressions such that the arguments of trigonometric functions come with positive sign.

\begin{figure*}[t]
\begin{center}
\includegraphics[width=\textwidth]{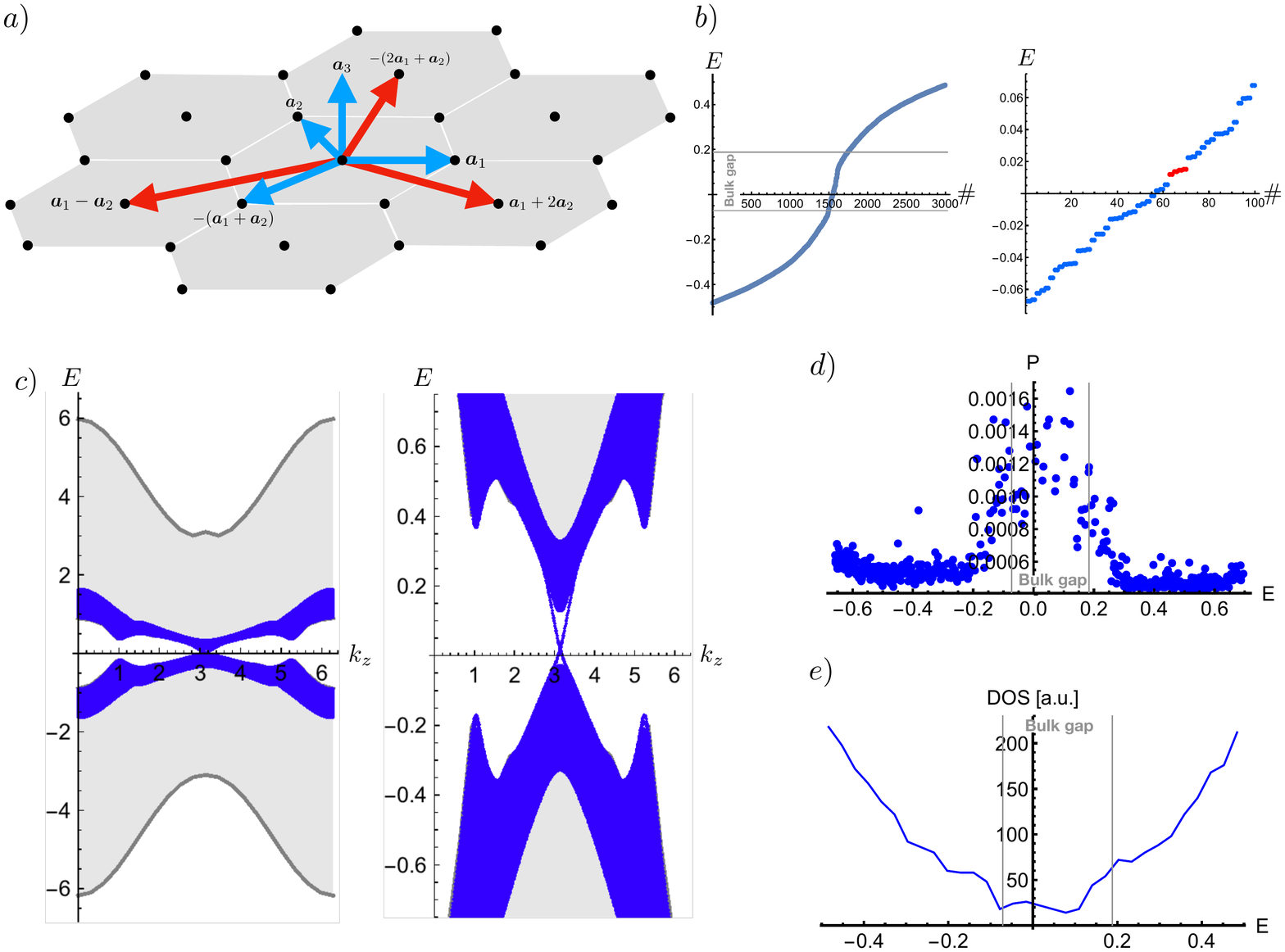}
\caption{
Tight-binding model for a HOTI with $\hat{C}_3$ and $\hat{I}$. 
a) 2D slice of the hexagonal crystal lattice. Nearest-neighbor and next-nearest-neighbor hoppings are indicated by blue and red arrows, respectively.
b) Spectrum of the Hamiltonian given in Eq.~\eqref{eq: TBHamiltonian} on a hexagonal geometry with open boundary conditions in all three directions. 3000 eigenstates are shown. Right panel: A closer view of 100 in-gap eigenstates. The sum of the real-space weights of the states in red is plotted in Fig.~\ref{fig: rod}~d) in the main text.
c) $k_z$-resolved spectrum of the tight-binding model with hexagonal cross-section and periodic boundary conditions in $z$-direction. The 1000 lowest lying eigenstates are shown in blue, the bulk spectrum in gray. The right panel represents a close-up on the low-energy part. The in-gap states at $k_z = \pi$ are twelve-fold degenerate, corresponding to six helical Kramers pairs of hinge modes. Consult Fig.~\ref{fig: rod}~f) in the main text for a further zoomed-in version.
d) Participation ratios of the 1000 eigenstates pertaining to the eigenvalues at $k_z = \pi$ in c). For an eigenstate $\ket{\Psi}$ with elements $\Psi_i = \braket{i | \Psi}$, $i = 1, \dots, N$ (where $N$ is the dimension of the full single-particle Hilbert space including lattice and on-site degrees of freedom), the participation ratio is defined as $\mathrm{P} = \sum_i |\Psi_i|^4$. It is small for bulk states and large for edge and corner states.
e) Density of states (DOS) corresponding to the spectrum in b). The one-dimensional hinge states have an approximately constant DOS and are visible around zero energy.
}
\label{fig: tightbinding}
\end{center}
\end{figure*}

The Hamiltonian consists of two 3D topological insulators, given by $H_\mathrm{TB, I}$ and $H_\mathrm{TB, II}$, which are coupled together via the mass matrix $M_\mathrm{TB}$. In the parameter regime $1 < |m_i| < 3$ for $i = \mathrm{I}, \mathrm{II}$, there is a single band inversion in the spectrum of $H_{\mathrm{TB},i}$ at $T$ and the respective Hamiltonian is in the topologically insulating phase. Outside of these two regimes, the full model given by $H_\mathrm{TB}$ is topologically trivial. In contrast, in the parameter regime where $1 < |m_i| < 3$ holds only for a single $i$, the full Hamiltonian represents a first-order topological insulator (assuming always that $\delta$ is small enough so as not to close the bulk gap), with a single Dirac cone on all surface terminations. Finally, in the regime where $1 < |m_i| < 3$ holds for all $i$, $H_\mathrm{TB}$ is trivial from a first order perspective, since it has an even number of band inversions, and indeed, for any small finite $\delta$, the two surface Dirac cones gap out. However, it is a HOTI, since $H_\mathrm{TB, \mathrm{I}}$ and $H_\mathrm{TB, \mathrm{II}}$ belong to different $C_3^z$ subspaces.

In all calculations we choose $m_\mathrm{I} = m_{\mathrm{II}} = 2$, $t_\mathrm{I} = t_{\mathrm{II}} = 1$, $\lambda_\mathrm{I} = 0.3$, $\lambda_{\mathrm{II}} = \gamma_{\mathrm{II}} = 1$, $\epsilon = 0.1$, and $\delta = 0.3$ to be deep in the HOTI phase. The particular values of these parameters are chosen so that the surface gaps of the tight-binding model in an open geometry are maximized in the HOTI phase. See Fig.~\ref{fig: tightbinding}~b) for the spectrum of the system in a hexagonal geometry cut out from a $30 \times 30 \times 20$ cuboid.

The HOTI Dirac theory presented in Supplementary Material section~\ref{sec: bulk-boundary}, as given by $H_\mathrm{D} (\bs{k})$ in Eq.~\eqref{eq: SIgappedDiracEquation}, is unitarily equivalent to an expansion of the present tight-binding model $H_\mathrm{TB}(\bs{k})$ about the $T$ point in the BZ.
The expansion is to linear order in $k_z$, to third order in $k_x$, $k_y$ for the terms resulting in the kinetic parts $H_1(\bs{k})$ and $H_3(\bs{k})$ and to second order in
 $k_x$, $k_y$ for the terms resulting in the mass-term. 
This expansion relates the parameters entering the two models as follows
\begin{equation}
\begin{aligned}
&M_0 = -3 t_\mathrm{I} = -3 t_{\mathrm{II}}, \quad M_1 = \frac{1}{4} M_0, \\
&\tilde{A} = \lambda_{\mathrm{I}} = \lambda_{\mathrm{II}} = - 3 \sqrt{3} \gamma_{\mathrm{II}}, \quad A_1 = \frac{3}{2} \tilde{A}, \quad A_3 = - \frac{1}{8} \tilde{A} \\
& a = \frac{3}{32} \left(5 \mathrm{i} - 3 \sqrt{3}\right) \delta, \quad b = \frac{3}{8} \left(5 + \sqrt{3} \mathrm{i}\right) \delta, \quad \mu = \epsilon.
\end{aligned}
\end{equation}
For this choice of parameters and to the orders prescribed above we then have that
\begin{equation}
U H_\mathrm{TB}(\bs{k}) U^\dagger \approx H_\mathrm{D} (\bs{k}), \quad U = \begin{pmatrix}
- e^{\mathrm{i} \frac{\pi}{4}} & 0 & 0 & 0 & 0 & 0 &0 &0 \\
0 & 0 & 0 & - \mathrm{i} e^{-\mathrm{i} \frac{\pi}{4}} & 0 & 0 &0 &0 \\
0 & \mathrm{i} e^{\mathrm{i} \frac{\pi}{4}} & 0 & 0 & 0 & 0 &0 &0 \\
0 & 0 & e^{-\mathrm{i} \frac{\pi}{4}} & 0 & 0 & 0 &0 &0 \\
0 & 0 & 0 & 0 & 0 & - e^{\mathrm{i} \frac{\pi}{4}} &0 &0 \\
0 & 0 & 0 & 0 & 0 & 0 &- \mathrm{i} e^{-\mathrm{i} \frac{\pi}{4}} &0 \\
0 & 0 & 0 & 0 & \mathrm{i} e^{\mathrm{i} \frac{\pi}{4}} & 0 &0 &0 \\
0 & 0 & 0 & 0 & 0 & 0 &0 &e^{-\mathrm{i} \frac{\pi}{4}}
\end{pmatrix}, \quad U U^\dagger = \mathbb{1}.
\end{equation}

\end{widetext}

\subsection{Irreducible representations for bismuth}
\label{sec: irreps}
Our goal is to tell if the three valence bands highlighted in red in Fig.~\ref{fig: rod}~e) in the main text can be expressed as any linear combination of pEBRs. 
The bismuth atoms are located at the Wyckoff position $6c$, which is a high-symmetry line that induces composite EBRs instead of pEBRs in momentum space. However, the site-symmetry group of $6c$ is a subgroup of the maximal Wyckoff positions $3a$ and $3b$, which induce pEBRs.
We thus consider the pEBRs induced from the Wyckoff positions $3a$ and $3b$ listed in Table~\ref{ebr166} as an example.
One can observe that, in order to have a pEBR, the 
number of occurrences of little group irreducible representations (irreps) at the different high symmetry points are not independent of one another. 
For instance, the pEBR $\bar \Gamma^{3a}_{4+5}\uparrow G$
contributes the irreps $\bar{X}_3\bar{X}_4$ to the valence bands at the $X$ point and the irreps $\bar{\Gamma}_4\bar{\Gamma}_5$ at the $\Gamma$ point. The pEBR $\bar \Gamma^{3a}_{8}\uparrow G$ also
contributes the irreps $\bar{X}_3\bar{X}_4$ to the valence bands at the $X$ point and the irrep $\bar{\Gamma}_8$ at the $\Gamma$ point.
From these considerations, we can deduce that the number of occurrences of $\bar{X}_3\bar{X}_4$ among all valence bands at the $X$ point needs to equal the number of occurrences of $\bar{\Gamma}_4\bar{\Gamma}_5$ and $\bar{\Gamma}_8$ at the $\Gamma$ point. 
We compare this constraint to the irreps computed for bismuth and listed in Table~\ref{irr-bi} and realize that it is not met: we find $\bar{X}_3\bar{X}_4$ once, but $\bar{\Gamma}_4\bar{\Gamma}_5$ and $\bar{\Gamma}_8$ together appear three times.

The same analysis has been applied to all \mbox{pEBRs} in the space group, taking into account all Wyckoff positions. 
The constraints arising in this general case are substantially more complex than in the above example and have been analyzed with a computer algorithm.
(The full information about high-symmetry momentum points can be found in the BCS). 
We find also in this general case that no linear combination of \mbox{pEBRs} matches the irreps in the valence bands of bismuth.
Therefore, bismuth must have nontrivial topology, in connection with a bulk band structure that is not expressible in terms of exponentially localized Wannier functions. Note in addition that the $\mathbb{Z}_4$ inversion index\cite{Khalaf17,SongZhang17} evaluates to $\kappa_1 = 2$ for bismuth, underlining its non-trivial topology from another point of view.

\begin{table}[h]
\caption{ pEBRs of space group 166 with TRS.  The notation $\bar \Gamma^{3a/b}_{j}\uparrow G$ stands for the induction of the site-symmetry group $\bar \Gamma^{3a/b}_{j}$ in the space group $G$, which subduces a pEBR.  \\ 
}
\begin{tabular}{c|cccc}
\hline
            & $X$ & $\Gamma$ & $L$ & $T$  \\
\hline
$3a$             &          &     &     &             \\
$\bar \Gamma^{3a}_{4+5}\uparrow G$ & $\bar{X}_3\bar{X}_4$ & $\bar{\Gamma}_4\bar{\Gamma}_5$ & $\bar{L}_3\bar{L}_4$ & $\bar{T}_4\bar{T}_5$   \\
$\bar \Gamma^{3a}_{6+7}\uparrow G$ & $\bar{X}_5\bar{X}_6$ & $\bar{\Gamma}_6\bar{\Gamma}_7$ & $\bar{L}_5\bar{L}_6$ & $\bar{T}_6\bar{T}_7$   \\
$\bar \Gamma^{3a}_{8  }\uparrow G$ & $\bar{X}_3\bar{X}_4$ & $\bar{\Gamma}_8$   & $\bar{L}_3\bar{L}_4$ & $\bar{T}_8$     \\
$\bar \Gamma^{3a}_{9  }\uparrow G$ & $\bar{X}_5\bar{X}_6$ & $\bar{\Gamma}_9$   & $\bar{L}_5\bar{L}_6$ & $\bar{T}_9$     \\
\hline
$3b$             &          &     &     & \\
$\bar \Gamma^{3b}_{4+5}\uparrow G$ & $\bar{X}_3\bar{X}_4$ & $\bar{\Gamma}_4\bar{\Gamma}_5$ & $\bar{L}_5\bar{L}_6$ & $\bar{T}_6\bar{T}_7$   \\
$\bar \Gamma^{3b}_{6+7}\uparrow G$ & $\bar{X}_5\bar{X}_6$ & $\bar{\Gamma}_6\bar{\Gamma}_7$ & $\bar{L}_3\bar{L}_4$ & $\bar{T}_4\bar{T}_5$   \\
$\bar \Gamma^{3b}_{8  }\uparrow G$ & $\bar{X}_3\bar{X}_4$ & $\bar{\Gamma}_8$   & $\bar{L}_5\bar{L}_6$ & $\bar{T}_9$     \\
$\bar \Gamma^{3b}_{9  }\uparrow G$ & $\bar{X}_5\bar{X}_6$ & $\bar{\Gamma}_9$   & $\bar{L}_3\bar{L}_4$ & $\bar{T}_8$     \\
\hline
\end{tabular}
\label{ebr166}
\end{table}

\begin{table}[h]
\caption{ 
The computed little-group irreps for bismuth are presented as follows.
}
\begin{tabular}{c|c}
        &  3 doubly-degenerate valence bands \\
\hline                
$X$ & 
$     \bar{X}_3\bar{X}_4 ;\
     \bar{X}_5\bar{X}_6 ;\
 \bar{X}_5\bar{X}_6 
$ \\                     
$  \Gamma   $ & 
$ 
     \bar{\Gamma}_8;\ \bar{\Gamma}_8;\ \bar{\Gamma}_4\bar{\Gamma}_5
$ \\
$  L   $ & 
$ 
     \bar{L}_3\bar{L}_4;\ \bar{L}_5\bar{L}_6;\  \bar{L}_5\bar{L}_6 
$ \\
$  T  $ & 
$ 
     \bar{T}_9;\ \bar{T}_8;\ \bar{T}_6\bar{T}_7 
$ \\
\hline                
\end{tabular}
\label{irr-bi}
\end{table}

\subsection{Effect of a band inversion at the $X$ points}
We will show here that, as claimed in the main text, adding a $\hat{C}_3$ symmetric triplet of negative inversion eigenvalue Kramers pairs to the occupied bands at $X_1$, $X_2$ and $X_3$ corresponds to the addition of a single such Kramers pair in the $e^{\mathrm{i} \pi}$ eigenvalue subspace of $\hat{C}_3$, and the addition of two in the $\mathrm{exp}(\pm\mathrm{i}\pi/3)$ subspace.

Let the added Kramers pair at $X_1$ be denoted by the eigenstates $\ket{u^k_{X_1}}$, $k=1,2$, where $\ket{u^2_{X_1}}=T \ket{u^1_{X_1}}$ and $T$ is the representation of the anti-unitary time-reversal operator. Take $C_3$ to be the representation matrix of threefold rotation. Since in a spinful system $\hat{C}_3^3 = -1$, we can always choose the gauge
\begin{equation}
C_3 \begin{pmatrix} \ket{u^k_{X_1}} \\
\ket{u^k_{X_2}}\\
\ket{u^k_{X_3}}
\end{pmatrix} = 
e^{\mathrm{i}\pi/3}
\begin{pmatrix}
0 & 1 & 0 \\
0 & 0 & 1 \\
1& 0 & 0
\end{pmatrix}
\begin{pmatrix} \ket{u^k_{X_1}} \\
\ket{u^k_{X_2}}\\
\ket{u^k_{X_3}}
\end{pmatrix}.
\end{equation}

This means that we can find eigenvectors
\begin{equation}
\begin{aligned}
\ket{u^k_{\pi}} &= e^{\mathrm{i} 2\pi/3} \ket{u^k_{X_1}} + e^{-\mathrm{i} 2\pi/3} \ket{u^k_{X_2}} + \ket{u^k_{X_3}}, \\
\ket{u^k_{+\pi/3}} &= \ket{u^k_{X_1}} + \ket{u^k_{X_2}} + \ket{u^k_{X_3}}, \\
\ket{u^k_{-\pi/3}} &= e^{-\mathrm{i} 2\pi/3} \ket{u^k_{X_1}} + e^{\mathrm{i} 2\pi/3} \ket{u^k_{X_2}} + \ket{u^k_{X_3}}.
\end{aligned}
\end{equation}
with eigenvalues $-1$, $e^{\mathrm{i} \pi/3}$, and $e^{- \mathrm{i} \pi/3}$, respectively. This means that when the added Kramers pairs are odd under inversion symmetry, the time-reversal invariant subspaces with $\hat{C}_3$ eigenvalues $-1$ and $e^{\pm \mathrm{i} \pi/3}$ are augmented by one and two such Kramers pairs, respectively. Being a $\mathbb{Z}_2$ quantity, only $\nu^{(\pi)}$ is affected by the inversion eigenvalues at the $X$ points. The same argument holds for the $L$ points.

\begin{figure}[t]
\centering
\includegraphics[width=0.48\textwidth]{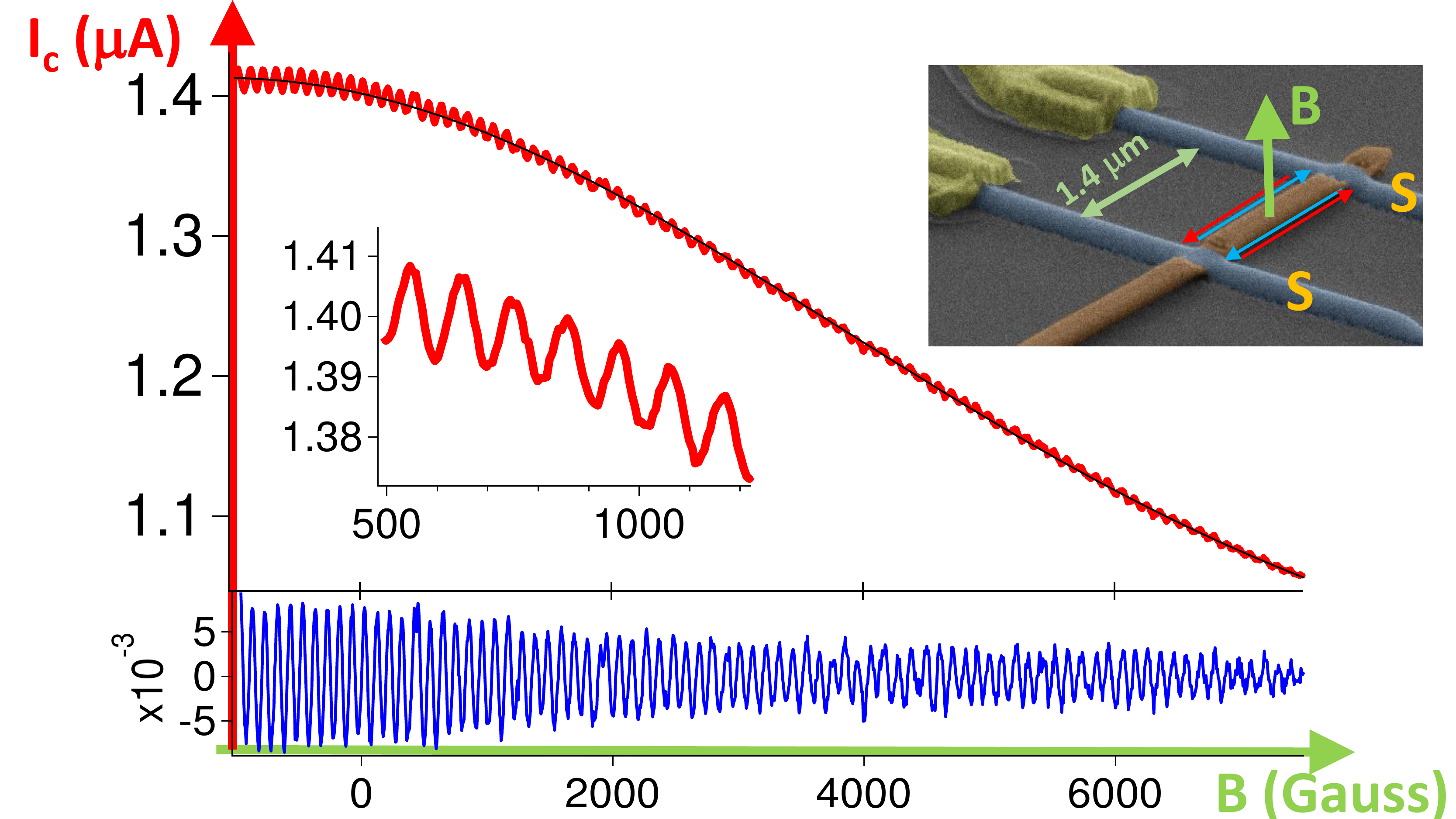}
\caption{
Field dependence of the Bi nanowire's critical current before insertion into the asymmetric SQUID configuration.
}
\label{fig: SI-transport}
\end{figure}

\subsection{Transport experiment: further details}
Superconducting electrodes were made by focused ion beam-assisted deposition of a superconducting tungsten compound. The $T_c$ and $H_c$ are above 4~K and 10~T, respectively. We show in Fig.~\ref{fig: SI-transport} the magnetic field dependence of the nanowire's critical current before insertion in the SQUID configuration: only two superconducting wires connect the Bi nanowire (there is no tungsten wire with constriction in parallel).
Periodic oscillations of the critical current with a period of 100 Gauss are visible (see zoomed-in figure). Since this periodicity corresponds to one magnetic flux quantum threading the wire (see Scanning Electron Micrograph with sample layout), it is the signature of interference between two supercurrent-carrying paths located  on two opposite edges of the wire. The small oscillation amplitude indicates an asymmetry in the transmission of the two paths.   
The decay of both these oscillations (bottom blue curve) and the total critical current, on a field scale of $B_1 = 8000$ Gauss,  gives an order of magnitude for the path width: $W=\Phi_0/L B_1 \simeq \SI{1}{\nano\meter}$. Whereas this simple measurement demonstrates the existence of two paths at the wire hinges, the supercurrent-versus-phase relation measurements described in the main text demonstrate the ballistic transport along those paths.

\end{document}